\documentclass[useAMS,usenatbib]{mn2e}
\usepackage{epsfig, bm, times, aas_macros}
\usepackage[english]{babel}
\usepackage{amsmath}
\usepackage{amssymb}
\usepackage{graphicx}
\usepackage{xcolor}

\newcommand{\seclab}{Section}
\newcommand{\tablab}{Table}

\newcommand{\figlab}{Figure}
\newcommand{\equlab}{Eq}
\newcommand{\secref}[1]{\seclab{} \ref{#1}}
\newcommand{\tabref}[1]{\tablab{} \ref{#1}}

\newcommand{\figref}[1]{\figlab{} \ref{#1}}
\newcommand{\figrefs}[1]{\figlab{s} \ref{#1}}
\newcommand{\eqr}[1]{\equlab{.} \eqref{#1}}

\newcommand{\timt}{\times}
\newcommand{\bc}{art-2012-braith-cavecchi}
\newcommand{\ca}{art-2013-cavecchi-etal}
\newcommand{\nuke}{\rm{n}}
\newcommand{\kaco}{\kappa_{\rm{c}}}
\newcommand{\tent}[1]{\times 10^{#1}}
\newcommand{\ro}{R_{\rm Ro}}

\newcommand{\crs}[1]{R_{\rm #1}}

\newcommand{\total}{\rm{D}  / \rm{D} t}
\newcommand{\tota}[1]{\frac{\rm{D} #1}{\rm{D} t}}
\newcommand{\phib}{\varphi}
\newcommand{\uthe}{{U^{\theta}}}
\newcommand{\uphi}{{U^{\phib}}}
\newcommand{\ur}{{U^{\rm r}}}
\newcommand{\nabsig}{{\bm\nabla}_\sigma}
\newcommand{\np}{N_{\rm p}}
\newcommand{\citeple}[1]{\citeauthor{#1}, \citeyear{#1}}
\newcommand{\x}{x}
\newcommand{\y}{y}
\newcommand{\z}{z}
\newcommand{\ux}{{U^{\rm \x}}}
\newcommand{\uy}{{U^{\rm \y}}}
\newcommand{\uz}{{U^{\rm \z}}}
\newcommand{\fr}{{F^{\rm r}}}
\newcommand{\fthe}{{F^{\theta}}}
\newcommand{\fphi}{{F^{\phib}}}

\newcommand{\fy}{{F^{\rm \y}}}
\newcommand{\fz}{{F^{\rm \z}}}
\newcommand{\fstar}{{F^{*}}}
\newcommand{\pot}{\tilde \phi}
\newcommand{\pota}{\phi}
\newcommand{\rstar}{\crs{S}}

\newcommand{\pstar}{P_\ast}
\newcommand{\ptop}{P_{\rm{top}}}
\newcommand{\pbot}{P_{\rm{bot}}}
\newcommand{\belt}{belt}
\newcommand{\tfive}{IGR~J17480-2446}

\title[Rotational effects in type I bursts]{Rotational effects in
  thermonuclear type I bursts: equatorial crossing and directionality
  of flame spreading}

\author[Y.\ Cavecchi et al.]{Yuri Cavecchi$^{1}$\thanks{E-mail:
    y.cavecchi@uva.nl}, Anna L.~Watts$^{1}$, Yuri Levin$^{2}$,
  Jonathan Braithwaite$^{3}$\\ $^{1}$Astronomical Institute ``Anton
  Pannekoek'', University of Amsterdam, Postbus 94249, 1090 GE
  Amsterdam, The Netherlands\\ $^{2}$Monash Center for Astrophysics
  and School of Physics, Monash University, Clayton, VIC 3800,
  Australia\\ $^{3}$Argelander Institut f\"ur Astronomie,
  Universit\"at Bonn, Auf dem H\"ugel 71, D-53121 Bonn, Germany}

\pagerange{\pageref{firstpage}--\pageref{lastpage}} \pubyear{}

\begin{document}
\maketitle
\label{firstpage}

\begin{abstract}
In a previous study on thermonuclear (type I) bursts on accreting
neutron stars we addressed and demonstrated the importance of the
effects of rotation, through the Coriolis force, on the propagation of
the burning flame. However, that study only analysed cases of
longitudinal propagation, where the Coriolis force coefficient
$2\Omega\cos\theta$ was constant. In this paper, we study the effects
of rotation on propagation in the meridional (latitudinal) direction,
where the Coriolis force changes from its maximum at the poles to zero
at the equator. We find that the zero Coriolis force at the equator,
while affecting the structure of the flame, does not prevent its
propagation from one hemisphere to another. We also observe structural
differences between the flame propagating towards the equator and that
propagating towards the pole, the second being faster. In the light of
the recent discovery of the low spin frequency of burster \tfive{}
rotating at 11 Hz (for which Coriolis effects should be negligible) we
also extend our simulations to slow rotation.
\end{abstract}


\section{Introduction}

Type I bursts are a phenomenon measured in more than 100 accreting
Neutron Stars (NSs) in low mass X-ray binaries (see MINBAR at
http://burst.sci.monash.edu/minbar, \citeple{rev-2010-gal-etal}).
During the bursts the matter accreted by the NS burns unstably after
reaching the critical column density \citep[see
  e.g.][]{art-1981-fuj-han-miy}. The time scales for different bursts
can vary, but in general they last from tens to hundreds of seconds
depending on different factors like the reactions taking place, the
ignition depth, the diffusion of photons through the non-burning
layers and the propagation of the flame across the surface
\citep[see for example][for reviews]{rev-1993-lew-par-taa,
  rev-2003-2006-stro-bild-book}.

Flame propagation during type I bursts is an essential component of
burst dynamics. Even if the accreted material is more or less
homogeneously distributed, it would be improbable that it would ignite
simultaneously everywhere on the surface \citep{art-1995-bild}. After
ignition is triggered at one location, flame propagation is important
to understand the observations of the rise of the lightcurve, but
there is more: NS parameters, like mass and radius, could be
derived from the lightcurves \citep[see, e.g.][for a recent
  review]{art-2010-guver-ozel-cabr-wrob,art-2010-stei-latt-brown,
  art-2011-sulei-pou-etal,art-2012-zam-cum-gall,
  rev-2013-miller-arxivpaper} and whether the flame is burning across
the full surface or only on one hemisphere, could have implications on
the inferred parameters.

Previously, \cite{art-2002-spit-levin-ush} employed analytical
arguments and shallow water numerical simulations to demonstrate the
defining role of rotation on the flame structure. Recently, \cite{\ca}
presented vertically resolved simulations of propagating deflagrations
in rotating oceans on the surface of NSs; they analysed the effects of
rotation by means of 2D numerical hydrodynamics simulations, using a
code described in \cite{\bc} and its modifications in \cite{\ca}. The
main conclusions of the latter paper were twofold. First, in the
presence of rotation the fluid is not free to move, so that, after
ignition, it expands vertically and tries to spill over sideways on to
the cold fluid. However, the Coriolis force prevents such motion,
diverting the fluid in the perpendicular direction and creating
hurricanes of fire that extend to two to three Rossby radii $\ro =
\sqrt(g H)/2\Omega$ (where $g$ is the gravitational acceleration at
the surface of the NS, $H$ the scale height of the fluid and $\Omega$
the angular velocity of the star). In this way the interface between
the hot and cold fluid, the flame front, where most of the burning is
happening, is along a line inclined at an angle $\sim H/\ro$: this is
in agreement with what \cite{art-2002-spit-levin-ush}
proposed. Secondly, inspection of the simulations revealed that the main
driver of the propagation, what makes the cold fluid ignite, is the
\emph{conduction} across the front (helped by fluid motion induced by
the baroclinicity at the hot-cold fluid interface). Therefore, the
speed is proportional to the thermal conductivity, $1/\kaco$, where
$\kaco$ is the opacity of the fluid. The timescales for conduction are
slow, but the front is not vertical, it is inclined as mentioned
above, therefore there is a geometrical factor, given by the inverse
of the inclination angle, of the order of $\ro/H$ that speeds up the
propagation bringing it to values comparable to observations
\citep{\ca}. This relation implied that the speed of the flame should
be proportional to the Rossby radius and therefore to $1/\Omega$.

However, \cite{\ca} considered only longitudinal propagation, using a
constant value of the Coriolis parameter for each run. Smaller values
of the spin exhibited faster propagation, but also less
confinement. This raised a very interesting question. Near the
equator, the Coriolis force vanishes and consequently so does the
confinement of the fluid. Could this prevent the propagation of the
fluid from one hemisphere to the other by quenching the flame? We
address this question in this paper.

For this work, while keeping the 2D setup, we have added the
possibility for the Coriolis parameter to vary with a cosine
dependence on the horizontal coordinate in order to mimic the
variation of the Coriolis force with latitude and analysed the
behaviour of the flame while approaching the equator. The structure of
this paper is the following: in the next section we report the results
of our simulations regarding meridional propagation and in
\secref{sec:conc} we draw our conclusions.

\section{Meridional propagation}
\label{sec:latprop}

In \cite{\ca} the question was raised whether a flame igniting
somewhere in one hemisphere could cross the equator and reach the
other hemisphere. The concern was that since the Coriolis force
vanishes at the equator, it would no longer be able to balance the
horizontal pressure gradient during the crossing, and the deflagration
would become de-confined and would fizzle out. This section is devoted
to the study of the flame propagation with non-constant Coriolis
parameter that goes to zero and switches sign in the middle of our
computational domain, but first we describe some necessary changes to
the code and the caveats.

\subsection{Equations of motions}
\label{sec:modi}

Cast in spherical coordinates, with $r$ along the outgoing radial
direction, $\phib$ increasing from west to east and $\theta$ going
from north to south, the north being defined by the positive direction
of the rotation axis, the equations of motion read:

\begin{align}
\tota{\ur} - \frac{\uthe^2 + \uphi^2}{r}=&
2\Omega \sin\theta \uphi - \frac{1}{\rho} \frac{\partial P}{\partial r} 
+ \frac{\partial \pot}{\partial r} + \frac{\fr}{\rho}\\
\tota{\uphi} + \frac{\ur \uphi}{r} + \frac{\uthe \uphi}{r \tan
  \theta}=& -2\Omega \cos\theta \uthe -2\Omega \sin\theta \ur +
\nonumber \\ 
& - \frac{1}{\rho}\frac{1}{r \sin \theta}\frac{\partial
  P}{\partial \phib} + \frac{1}{r \sin \theta}\frac{\partial \pot}{\partial
  \phib} + \frac{\fphi}{\rho}\\
\tota{\uthe} + \frac{\ur \uthe}{r} - \frac{\uphi^2}{r \tan \theta}=&
2\Omega \cos\theta \uphi - 
\frac{1}{\rho} \frac{1}{r}\frac{\partial P}{\partial \theta} 
+ \frac{1}{r}\frac{\partial \pot}{\partial \theta} + \frac{\fthe}{\rho}\\
\tota{\rho} = -\rho\left\{\frac{\partial \ur}{\partial r} +
\frac{1}{\rho}\frac{1}{r \sin \theta}\right.&\left.
\frac{\partial \uphi}{\partial
  \phib}+ \frac{1}{r}\frac{\partial \uthe}{\partial \theta} +
\frac{2 \ur}{r} + \frac{\uthe}{r \tan \theta} \right\}
\end{align}

\noindent Where $\total$ expresses the total derivative
\begin{equation}
\tota{} = \frac{\partial}{\partial t} + \ur \frac{\partial}{\partial r }
+  \uphi\frac{1}{r \sin \theta}\frac{ \partial}{\partial \phib} +
\uthe  \frac{1}{r}\frac{ \partial}{\partial \theta},
\end{equation} 
and the three velocities are $\ur = \dot{r}$, $\uphi =
r\sin\theta\dot{\phib}$, $\uthe = r \dot{\theta}$. $\Omega$ is the
angular velocity, and $\pot = -g r + \Omega^2r^2\sin^2\theta/2 $ is the
potential corrected for the centrifugal forces. $\fstar$ are the
viscous forces per unit volume.

The form of $\pot$ clearly shows that the coordinate surfaces
$r=\rm{const}$ are not potential surfaces anymore and this results in
a component of the potential force in the horizontal
direction. However, as is customary in geophysical sciences
\citep{art-2005-white-hosk-roul-stani}, if we are interested in a thin
layer whose mass is not contributing significantly to the
gravitational force, we can approximate the potential surfaces to
spheres and also drop any curvature term not proportional to
$1/\tan\theta$. In doing so, we set $r = \rstar + \x$, with $\rstar
\gg \x$, so that we can approximate $r$ to $\rstar$, apart from where
differentiation is involved, and $\pot = -g \x$, where we safely
remove the constant $-g \rstar$ from the potential \citep[the
  dependence of the potential on the colatitude should not be present,
  otherwise fake vorticity is introduced,
  see][]{art-2005-white-hosk-roul-stani}; finally, we can reduce the
$r$ component of the momentum equation to the hydrostatic equilibrium
one. In summary we have:

\begin{align}
\label{eq:approxux}
\frac{1}{\rho} \frac{\partial P}{\partial \x} =&
\frac{\partial \pot}{\partial \x}\\
\tota{\uphi} + \frac{\uthe \uphi}{\rstar \tan \theta}=&
-2\Omega \cos\theta \uthe  - 
\frac{1}{\rho}\frac{1}{\rstar \sin \theta}\frac{\partial P}{\partial \phib} 
+ \nonumber \\
\label{eq:approxuy}
&+ \frac{1}{\rstar \sin \theta}\frac{\partial \pot}{\partial \phib} + 
\frac{\fphi}{\rho}\\
\label{eq:approxuz}
\tota{\uthe} - \frac{\uphi^2}{\rstar \tan \theta}=&
2\Omega \cos\theta \uphi - 
\frac{1}{\rho} \frac{1}{\rstar}\frac{\partial P}{\partial \theta}
+  \frac{1}{\rstar}\frac{\partial \pot}{\partial \theta} 
+ \frac{\fthe}{\rho}\\
\label{eq:approxrho}
\tota{\rho} = -\rho\left\{\frac{\partial \ur}{\partial \x} 
+\right.&\left.  \frac{1}{\rstar \sin \theta}
\frac{\partial \uphi}{\partial \phib}+
 \frac{1}{\rstar} \frac{\partial \uthe}{\partial \theta} 
+ \frac{\uthe}{\rstar \tan \theta}
\right\}
\end{align}
with
\begin{equation}
\tota{} = \frac{\partial}{\partial t} + \ur \frac{ \partial}{\partial
  \x} + \uphi\frac{1}{\rstar \sin \theta} \frac{ \partial}{\partial \phib} +
\uthe \frac{1}{\rstar}\frac{ \partial}{\partial \theta}
\end{equation} 
The error in these expressions is then of order the eccentricity
squared $e^2$ or the angular velocity measured in units of the
(Newtonian) break-up velocity $\mu$ \cite[$e=\sqrt{a^2-b^2}/a$,
  $\mu=a^3\Omega^2/G/M$, where $a$ and $b$ are the major and minor
  axis of the ellipsoid which would better approximate the
  star]{art-2008-toorn-zimmerm}. An approximate estimation of $e$ and
$\mu$ can be derived following \citet{art-2007-mor-lea-cad-brag}. They
present interpolating formulae to estimate $b$ (their $R(0)$) given
the mass $M$, the radius $\rstar$ (their $R_{\rm eq}$) and
$\Omega=2\pi\nu$. In this study, we use $M=1.4 \rm{M_{\odot}}$ and
$\rstar=(30/\pi) 10^5$ cm; the spin frequency $\nu$ is at most $10^3$
Hz: in this case the maximum value for both $e$ and $\mu$ is $\sim
0.2$. Our reference case has $\nu=450$ Hz, with $e$ and $\mu$ $\sim
0.04$, resulting in an error of at most a few percent.

We use the hydrostatic numerical scheme described in \cite{\bc}, as
modified in \cite{\ca}, and further include the terms that account for
the spherical geometry. To do so, we change the vertical coordinate
from $\x$ to pressure $\sigma$. This is a hybrid coordinate system
that relies on the hydrostatic approximation. $\sigma$ is defined in
the following way: $P=\sigma \pstar + \ptop$ with $\pstar = \pbot -
\ptop$ \citep[see][]{art-1974-kasa,\bc}. In this coordinate system the
upper boundary is fixed in pressure and the lower in space. $\pstar$
is the pressure difference between bottom and top and changes as the
fluid moves around. This system is well suited for NS oceans where the
domain is much more extended horizontally than vertically and the
hydrostatic assumption is justified; it also allows the grid to follow
the vertical expansion of the fluid without the need for excessive
memory \citep[see][for a more detailed discussion]{\bc}. After setting
$\rstar \sin\theta {\rm{d}} \phib = {\rm{d}} \y$, $\rstar {\rm{d}} \theta =
{\rm{d}} \z$, $\ur = \ux$, $\uphi = -\uy$, $\uthe = \uz$ and
$\pota=-\pot=g\x$ for convenience, our equations read

\begin{align}
\label{eq:flatux}
 \frac{\partial \sigma}{\partial \x} =&
-\frac{g\rho}{\pstar}\\
\label{eq:flatuy}
\tota{\uy} + \frac{\uz \uy}{\rstar \tan \theta}=&
2\Omega \cos\theta \uz 
- \frac{\sigma}{\rho}\frac{\partial \pstar}{\partial \y} 
- \frac{\partial \pota}{\partial \y} + \frac{\fy}{\rho}\\
\label{eq:flatuz}
\tota{\uz} - \frac{\uy^2}{\rstar \tan \theta}=&
-2\Omega \cos\theta \uy - \frac{\sigma}{\rho} \frac{\partial \pstar}{\partial \z} 
- \frac{\partial \pota}{\partial \z} + \frac{\fz}{\rho}\\
\label{eq:flatrho}
 \tota{\rho} + \rho\left\{\frac{\partial
  \dot\sigma}{\partial \sigma} + \frac{\partial \uy}{\partial \y}\right.
+&\left. \frac{\partial \uz}{\partial \z} + \frac{\uz}{\rstar \tan \theta}
\right\} = 0
\end{align}
with
\begin{equation}
\tota{} = \frac{\partial }{\partial t} + \dot\sigma \frac{
  \partial}{\partial \sigma} + \uy \frac{ \partial}{\partial \y} +
\uz\frac{ \partial}{\partial \z}
\end{equation} 
and
\begin{equation}
\nabsig  = \frac{ \partial }{\partial \y} +
\frac{ \partial }{\partial \z}
\end{equation} 

\noindent Continuity \eqr{eq:flatrho} changes into an equation for $\pstar$

\begin{equation}
\label{eq:flatpstar}
\frac{\partial \pstar}{\partial t} = -I_{\sigma=1}
\end{equation}
with
\begin{equation}
I\equiv
\int^\sigma_0\!\!\left({\bm\nabla}_\sigma \!\cdot\!(\pstar{\bf U}) +  
\frac{\pstar \uz}{\rstar \tan \theta}\right)\,{\rm  d}\sigma'
\end{equation}
and
\begin{equation}
\dot\sigma  = \frac{1}{\pstar}
\left( \sigma I_{\sigma=1} - I \right)
\end{equation} 
just as in \cite{\bc}, but with the new definition of $I$ with the
extra term in $1/\tan\theta$ due to the use of spherical coordinates.

As for the energy equation, equation 11 in \cite{\ca}, it involves
only scalar terms and they do not change, apart from the total
derivative and the conduction term. As for this latter (\equlab 28 of
\citeple{\ca}) a similar treatment as above \citep[see equation 3.17 of][
  translating $a=\rstar$ and $\phi=\pi/2 -
  \theta$]{art-2005-white-hosk-roul-stani} leads to

\begin{multline}
\label{eq:flatTemp}
Q_{\rm{cond}} = 
\frac{1}{P_\star}\left\{{\bm\nabla}_\sigma
  \left[
    \frac{16\sigma_BT^3}{3\kaco\rho}
    \left(
      \frac{P_\star}{\rho}{\bm\nabla}_\sigma T + 
      {\bm\nabla}_\sigma\phi\frac{\partial T}{\partial \sigma}
    \right)
  \right]+
\right.\\
+\frac{\partial}{\partial\sigma}
\left[
  \frac{16\sigma_BT^3}{3\kaco P_\star}
  \left(
    {\bm\nabla}_\sigma \phi^2\frac{\partial T}{\partial \sigma} 
    +\frac{P_\star}{\rho}
    {\bm\nabla}_\sigma \phi {\bm\nabla}_\sigma T
  \right.+
\right.\\
\left.
  \left.
    \left.      
      + g^2\frac{\partial T}{\partial\sigma}
    \right)
  \right]
+ \frac{16\sigma_BT^3}{3\kaco\rho\tan\theta}\left(
\frac{P_\star}{\rho}\frac{\partial T}{\partial \z} + 
\frac{\partial \phi}{\partial \z}\frac{\partial T}{\partial \sigma}
\right)\right\}
\end{multline}
where $\sigma_{\rm B}$ is the Stefan--Boltzmann constant and $\kaco$ is the
opacity, our parametrization of thermal conductivity, whose importance
as the leading mechanism responsible for flame propagation was
demonstrated in \cite{\ca}. 

Thus, all our equations of motion look like those of \cite{\bc} and
\cite{\ca}. The only essential differences are the cosine dependence
of the Coriolis force and the terms in $1/\tan\theta$. These we have
implemented to simulate the variation of the Coriolis parameter from pole
to pole and the effects of curvature\footnote{The terms in
  $1/\tan\theta$ diverge at the poles, but the symmetries we impose on
  our problems, i.e. reflective boundary conditions as in \cite{\ca},
  make the terms go to zero in such loci.}.

A final remark regards the derivatives in the `$y$' direction: this is
the longitudinal direction and it is clear that dealing with it
requires particular care, since we are dealing with circles of
different length.  As a first approximation we assume symmetry along
the longitudinal direction, therefore setting to zero every derivative
along $y$. The symmetry of our problems allows for such
simplification: when igniting at the pole every longitude should be
treated equally, while when igniting at the equator we are implying a
ring ignition.

\subsection{Initial setup}
\label{sec:numset}

\begin{table}
  \begin{center}
    \begin{tabular}{|c|l|r|c|}
      \hline
      Name & $\kaco$ [cm$^{2}$ g$^{-1}$] & $\nu$ [Hz] & $\delta$ [$10^5$ cm]\\
      \hline
      P1 / E1 & $7\tent{-2}$ & $  10$&   ---   \\
      P2 / E2 & $7\tent{-2}$ & $ 100$&  $6.03$ \\
      P3 / E3 & $7\tent{-2}$ & $ 450$&  $2.77$ \\
      P4 / E4 & $7\tent{-2}$ & $1000$&  $1.85$ \\
      P5 / E5 & $7         $ & $ 450$&  $2.77$ \\
      P6 / E6 & $7\tent{-1}$ & $ 450$&  $2.77$ \\
      P7 / E7 & $7\tent{-3}$ & $ 450$&  $2.77$ \\
      \hline
    \end{tabular}
    \caption{Numerical parameters for the various simulations. $\kaco$
      is the opacity, $\nu$ is the spin frequency of the
      star. $\delta$ is the theoretical half width of the equatorial
      \belt{} where the Coriolis force is not capable of confining the
      fluid. Simulations with $\nu = 10$ Hz are not confined. All
      simulations have vertical resolution of $m_{\rm{\x}} = 96$. The
      horizontal resolution is $m_{\rm{\z}} = 480$ for the setups igniting at
      the pole (P) and it is $m_{\rm{\z}} = 240$ for those igniting at the
      equator (E).}
    \label{tab:poleq}
  \end{center}
\end{table}

We set up our simulations in a similar fashion to \cite{\ca}. In
particular, the fluid is initially at rest and is made of pure helium,
burning into carbon according to
\begin{equation}
Q_{\nuke}=5.3\timt10^{18}\rho_5^2 \left(\frac{Y}{T_9}\right)^3 
e^{-4.4/T_9}\; \textrm{erg g$^{-1}$ s$^{-1}$},
\end{equation}
where $T_9$ is temperature in units of $10^9$ K, $\rho_5$ is density
in units of $10^5$ g cm$^{-3}$ and $Y$ is the mass fraction of
helium. The temperature distribution is vertically constant with a
horizontal dependence as:
\begin{equation}
T = T_0 + \frac{\delta T}{1 + \exp[(\z - 0.9\;\rm{km})/0.36\;\rm{km}]}
\end{equation}
so that there is a greater temperature at one end of the domain,
designed to ignite the fluid, whilst the distribution is flat in the
rest of the domain, see \cite{\ca} for details. Here we use
$T_0=2\tent{8}$ K and $\delta T=2.81\tent{8}$ K. The lower
  value, $T_0=2\tent{8}$ K, is chosen to be neither too low, since the
  whole ocean must be close to ignition, nor so high as to trigger
  self ignition \citep[see the ocean temperature profiles of][but note
    that the rate of energy production of the triple $\alpha$
    reactions used by the latter authors is 1.9 times the rate used in
    this paper, since the authors apply this approximation to take
    carbon burning into account]{art-1995-bild,art-2000-cum-bild}.
The opacity $\kaco$ is constant in each simulation, but different for
different simulations (see \tabref{tab:poleq} of this paper and
  the discussion at the end of section 2.2 of \citeauthor{\ca},
  \citeyear{\ca}). The stellar spin $\nu$ varies between simulations
from $10$ to $10^{3}$ Hz (see \tabref{tab:poleq}), while the
surface gravity is $g=2\tent{14}$ cm s$^{-2}$.  The horizontal length
of the domain, the hemicircumference of the star, is $3\tent{6}$ cm,
corresponding to $\rstar=(30 / \pi)$ km, while in the vertical
direction we include layers from $\ptop=10^{22}$ Pa to
$P_{\rm{bot}} = e^{1.7}\tent{22}$ Pa. In the simulations for polar ignition the
cosine term in the Coriolis force goes from $1$ to $-1$ across the
domain. In the simulations for equatorial ignition, it goes from $0$
to $-1$.

The horizontal boundaries are symmetric in pressure difference
  $P_\ast$, temperature $T$ and composition. They are antisymmetric in
  the horizontal velocities. At the upper and lower boundaries, we use
  symmetric conditions for the temperature, composition and horizontal
  velocities. However, we include a cooling term, that affects the top
  of the simulation, based on an approximation of heat losses
  \citep[see][]{\ca}. We do not include any heat flux from the bottom
  boundary. In the case of equatorial ignition, the symmetry allows us
  to simulate just one hemisphere; simulations with polar ignition
  model the entire surface. We assume axisymmetry in both cases, thus
  limiting the simulations to 2D. The simulations have the same
vertical resolution $m_{\rm{\x}}=96$ grid points and horizontal
resolutions of $m_{\rm{\z}}=480$ for those that ignite at the pole and
$m_{\rm{\z}}=240$ for those igniting at the equator: that is
equivalent to $6250$ cm in both cases. Artificial diffusivities are
$\nu_1=0.03$ and $\nu_2=0.5$ \citep[see][for a description of the
  diffusion schemes implemented]{\bc}.

\subsection{Polar ignition and equatorial crossing}
\label{sec:poligni}

Our key goal was to find out whether flames always crossed the
equator, or whether the loss of Coriolis confinement led to
quenching. We found that in every case we studied the fluid was
eventually burning over the whole star. The simulations with the
slowest rotation, $\nu = 10$ Hz, were qualitatively different since
the Rossby radius is larger than the star and no effective confinement
is ever realized: this is discussed in more detail below (see
\secref{sec:norot}). On the other hand, all the other runs developed a
well defined flame front that crossed the equator in all cases.

\figref{fig:crossP3a} shows the crossing of the equator for our
reference run P3\footnote{\label{note:units}Note that in
    \citet{\ca} the figures for the heating rates were erroneously
    labelled with units K/s rather than erg/g/s, omitting the factor
    of the gas constant.}. The flame proceeds from the pole to the
equator with a configuration similar to that described in \cite{\ca}
(see \figref{fig:crossP3b}, panel at $\approx 3.53$ s), the difference
being a decreasing Coriolis force and increasing $\ro$. This manifests
itself as a flame front which is increasingly close to being
horizontal and a speed up of flame propagation. When the flame is near
enough to the equator, the Coriolis force is no longer able to
significantly confine the hot fluid and this starts spilling over the
cold fluid around the equator; the fluid is stopped in the Southern
hemisphere once the Coriolis force is significant again (panel at
$\approx 4.42$ s). As the burning layer on the other side is heating
the layer below, the flame front is nearly horizontal. Thus, in the
equatorial region the flame propagates vertically downwards (panels at
$\approx 4.68$ - $\approx 4.94$ s). After consuming the equatorial
\belt{} the flame propagates in the Southern hemisphere (last panel of
\figref{fig:crossP3b}, see also \secref{sec:propfit}).

It is useful to define the equatorial \belt{} as the region where the
Coriolis force plays no essential dynamical role. Its extent is
limited by the latitude on both sides of the equator where the distance
to the equator is equal to the Rossby radius $\ro = \sqrt{g H} / 2
\Omega \cos\theta$ at that point (and symmetrically on the other
side):

\begin{equation}
\label{eq:belt}
\z - \frac{\pi}{2}\rstar = \frac{\sqrt{g H}}{2\Omega\cos(\z/\rstar)}
\end{equation}

The solutions to \eqr{eq:belt} can be transformed to the \belt{} half
width, $\delta = \z - \frac{\pi}{2}\rstar$. Theoretical \belt{} widths
for each simulation are reported in \tabref{tab:poleq} and drawn on
\figrefs{fig:crossP3a} - \ref{fig:crossP7} and \ref{fig:propP3} for
comparison with the results. Simulations with $\nu = 10$ Hz do not
have a \belt{} width, in the sense that \eqr{eq:belt} does not have a
meaningful solution since the Rossby radius is bigger than the star.

\begin{figure*}
  \centering
  \includegraphics[width=0.45\textwidth]{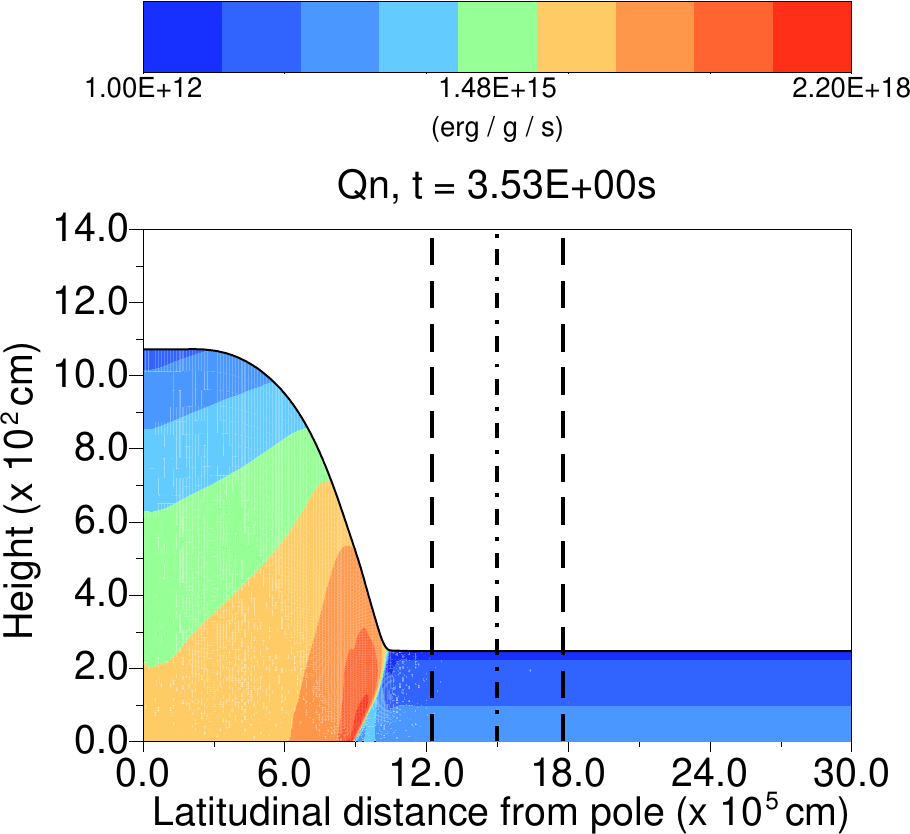}a)
  \hspace{\stretch{1}}
  \includegraphics[width=0.45\textwidth]{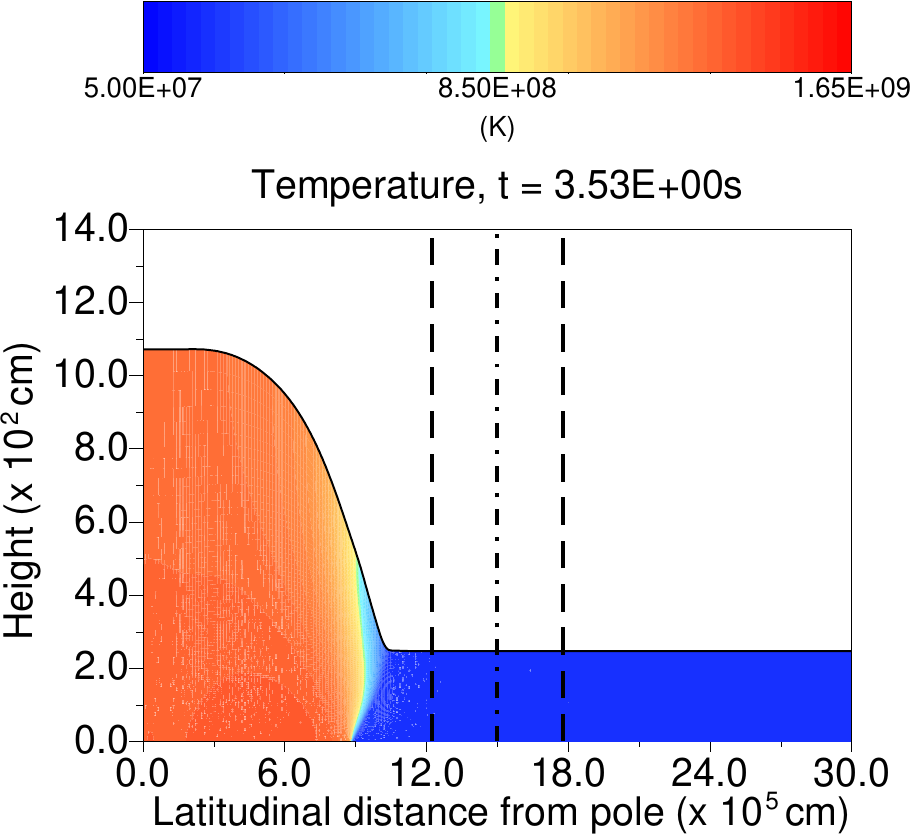}b)\\
  \includegraphics[width=0.45\textwidth]{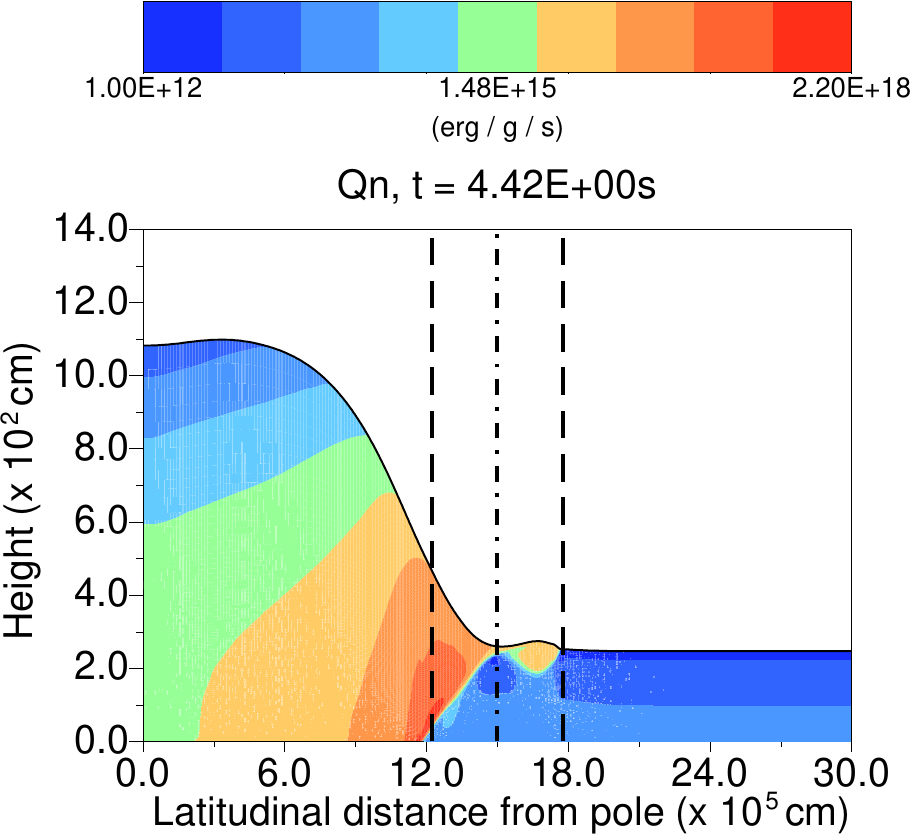}c)
  \hspace{\stretch{1}}
  \includegraphics[width=0.45\textwidth]{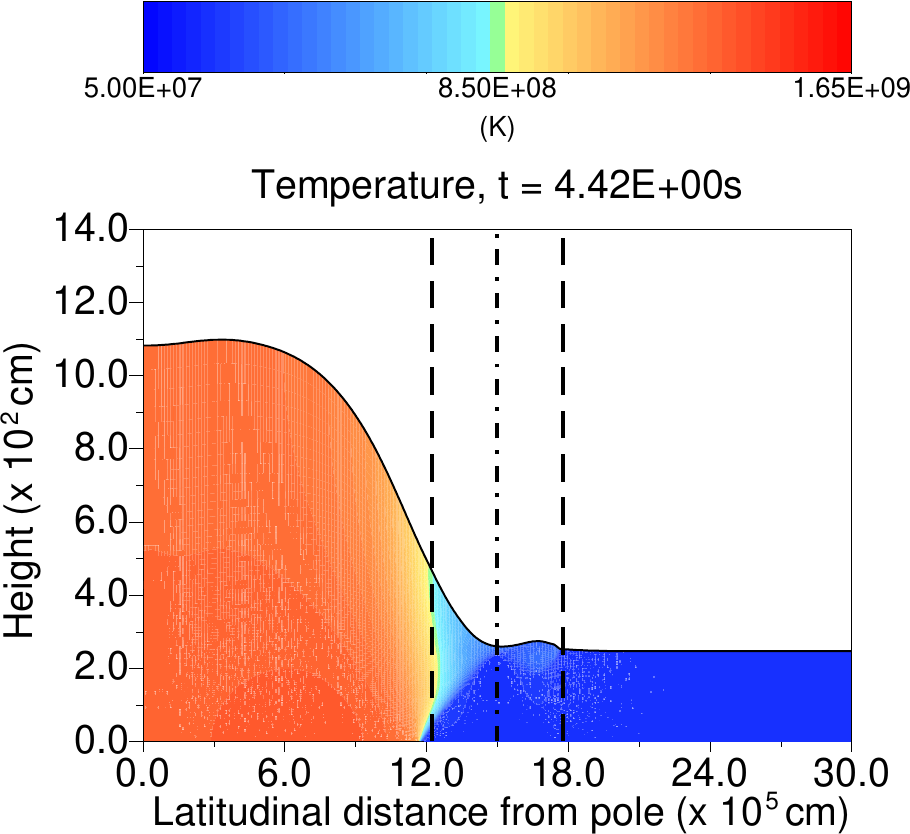}d)\\
  \includegraphics[width=0.45\textwidth]{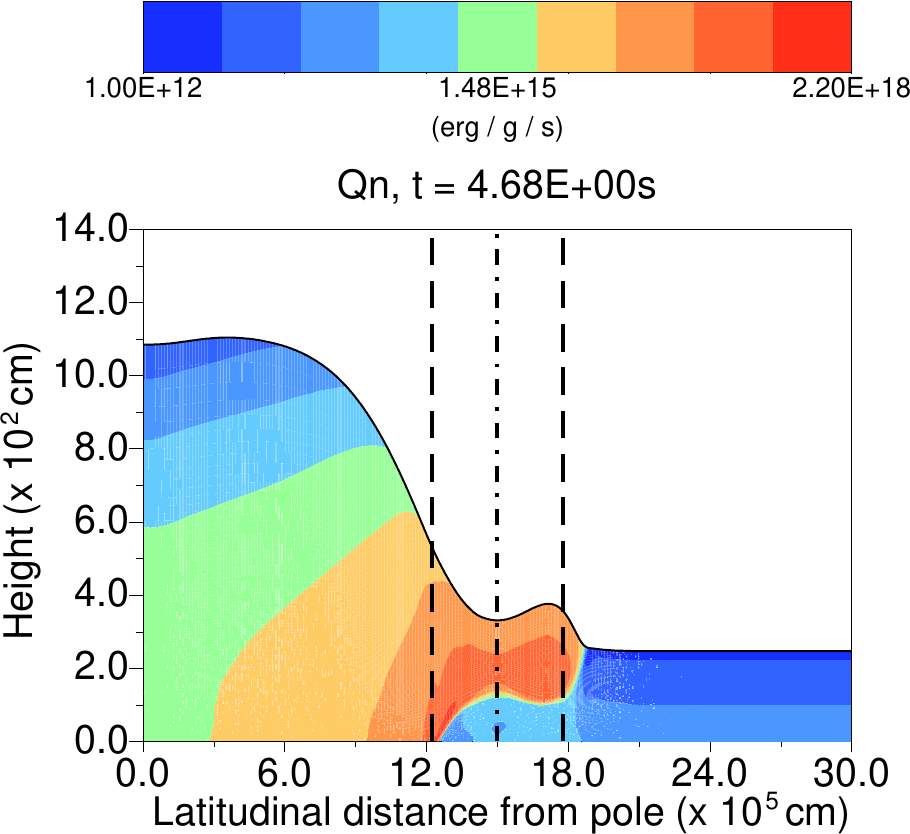}e)
  \hspace{\stretch{1}}
  \includegraphics[width=0.45\textwidth]{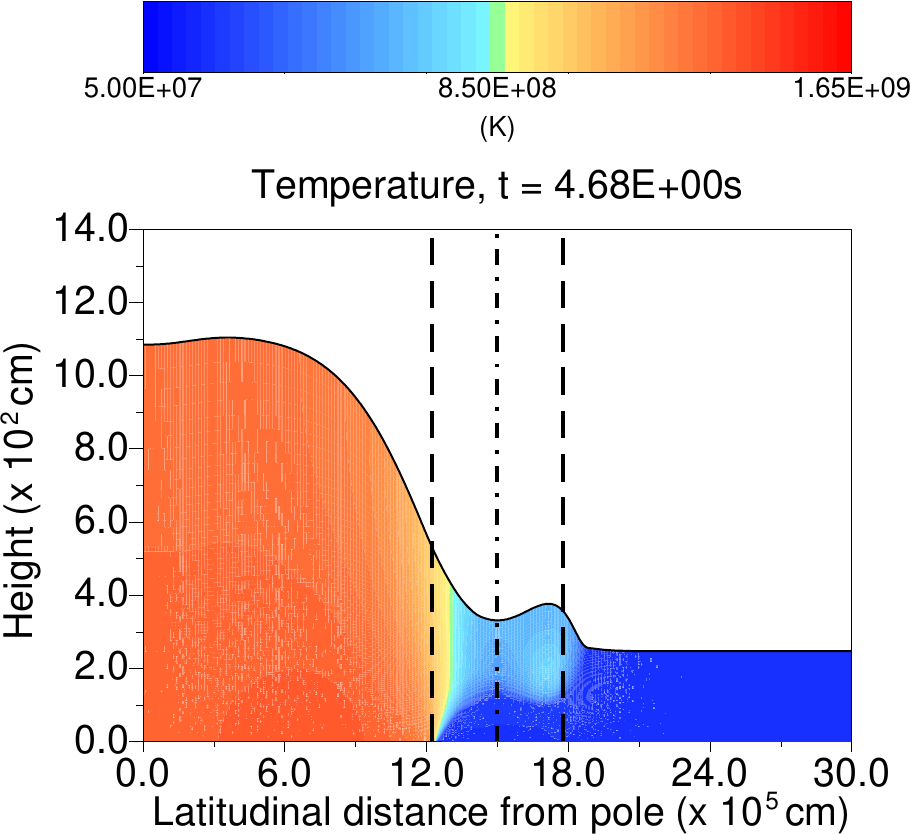}f)\\
  \caption{The crossing of equator of P3 ($\nu=450$ Hz,
    $\kaco=7\tent{-2}$ cm$^{2}$ g$^{-1}$). Heating rate due to
      nuclear burning, $Q_{\rm n}/{\tilde c_{\rm P}}$ as in equation
      23 of \citeauthor{\ca}, (\citeyear{\ca}), scaled by the gas
      constant value to make it in erg g$^{-1}$ s$^{-1}$ (left-hand column,
      logarithmic scale)$^{\ref{note:units}}$ and temperature (right-hand
      column, linear scale). Vertical lines indicate the equator
    (dash dotted) and the \belt{} (dashed, see \equlab{}
    \ref{eq:belt}). The flame propagates confined by the Coriolis
    force in the Northern hemisphere. When it reaches the \belt{},
    confinement is not enough and the hot fluid begins spilling over
    the cold one. At the southern extreme of the \belt{} Coriolis
    force is effective again and the fluid is confined again.}
  \label{fig:crossP3a}
\end{figure*}

\addtocounter{figure}{-1}
\begin{figure*}
  \centering
  \includegraphics[width=0.45\textwidth]{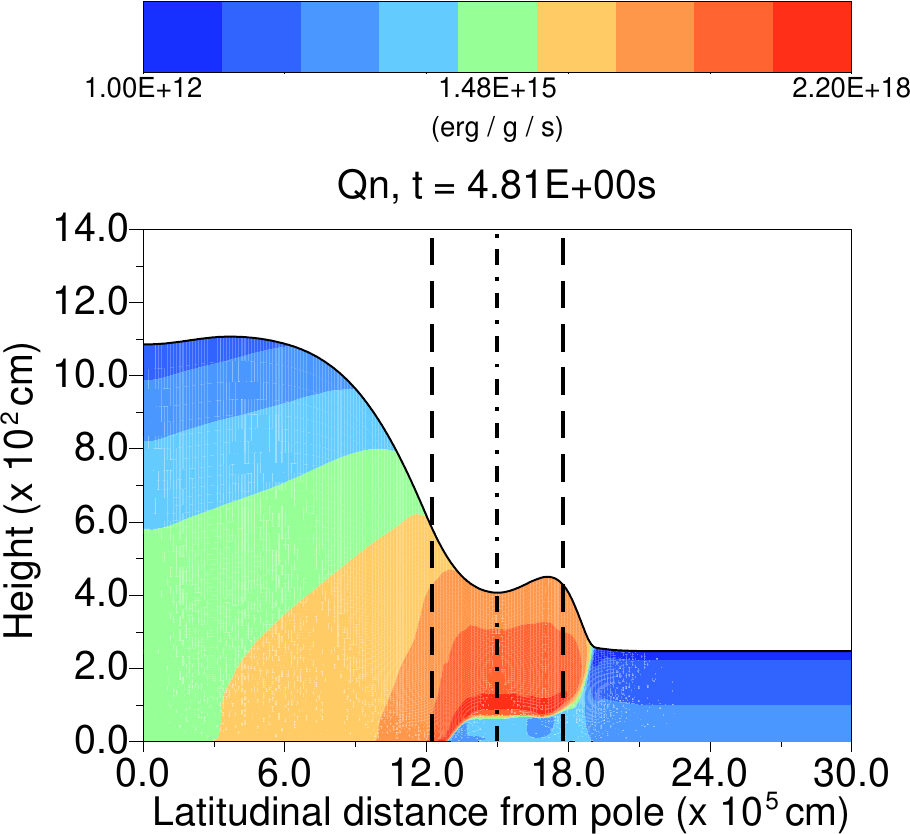}g)
  \hspace{\stretch{1}}
  \includegraphics[width=0.45\textwidth]{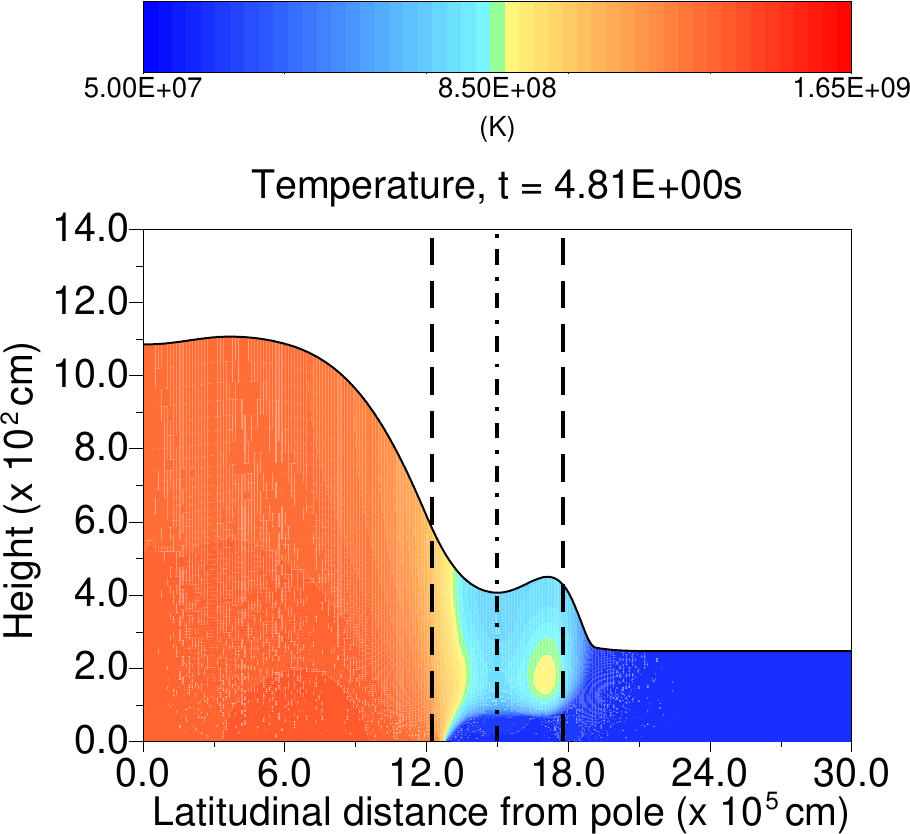}h)\\
  \includegraphics[width=0.45\textwidth]{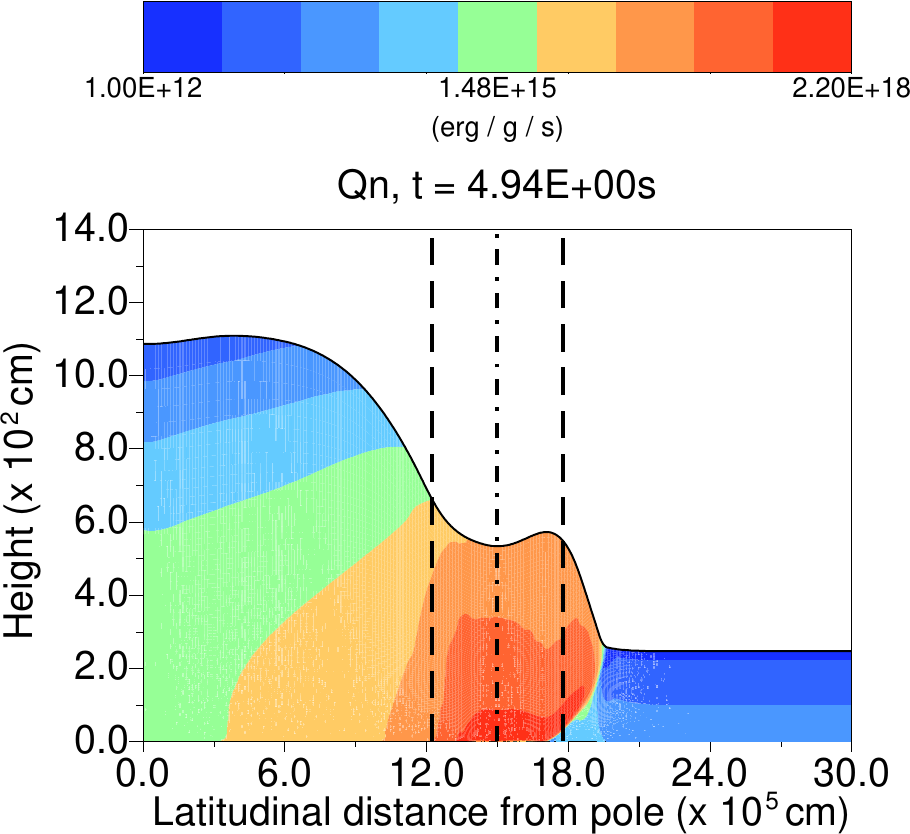}i)
  \hspace{\stretch{1}}
  \includegraphics[width=0.45\textwidth]{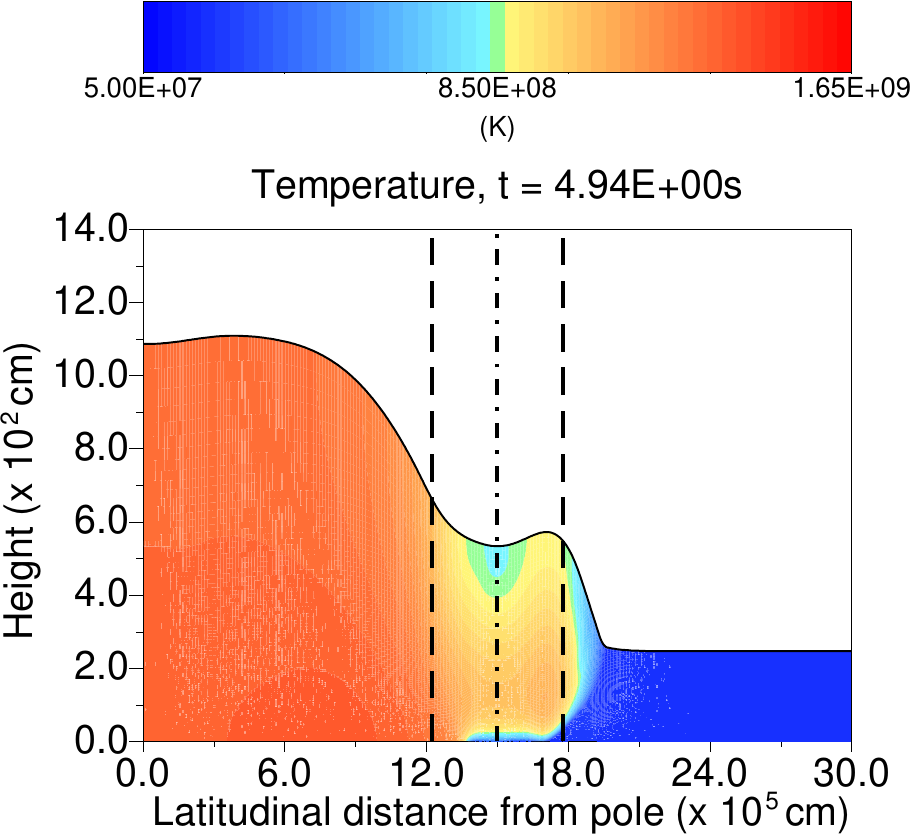}j)\\
  \includegraphics[width=0.45\textwidth]{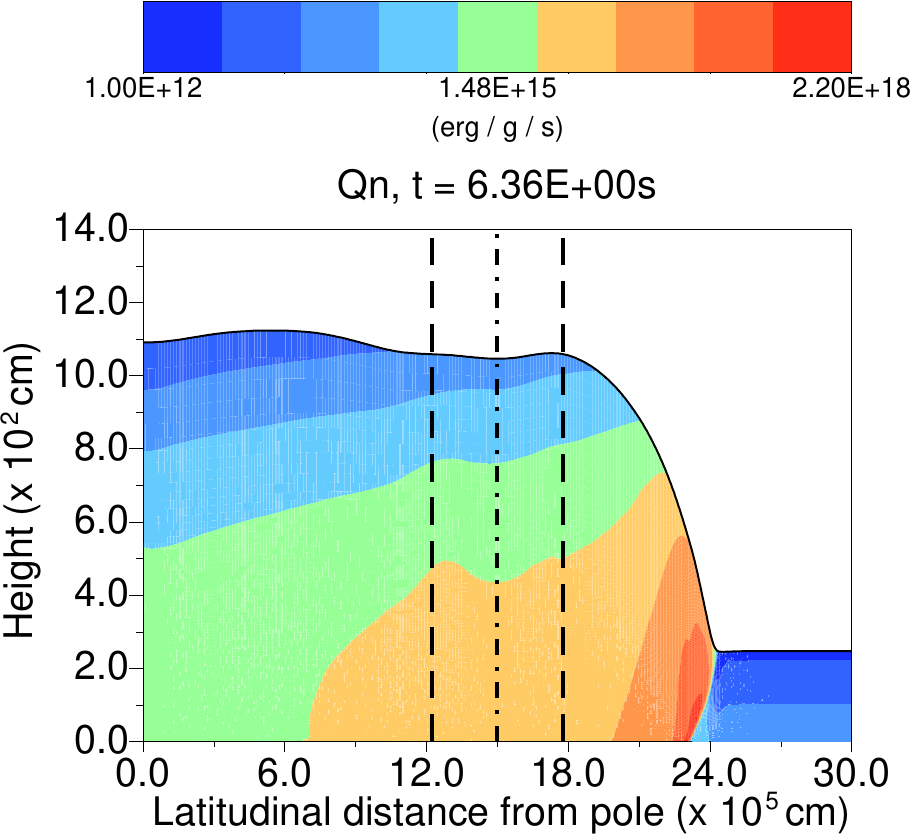}k)
  \hspace{\stretch{1}}
  \includegraphics[width=0.45\textwidth]{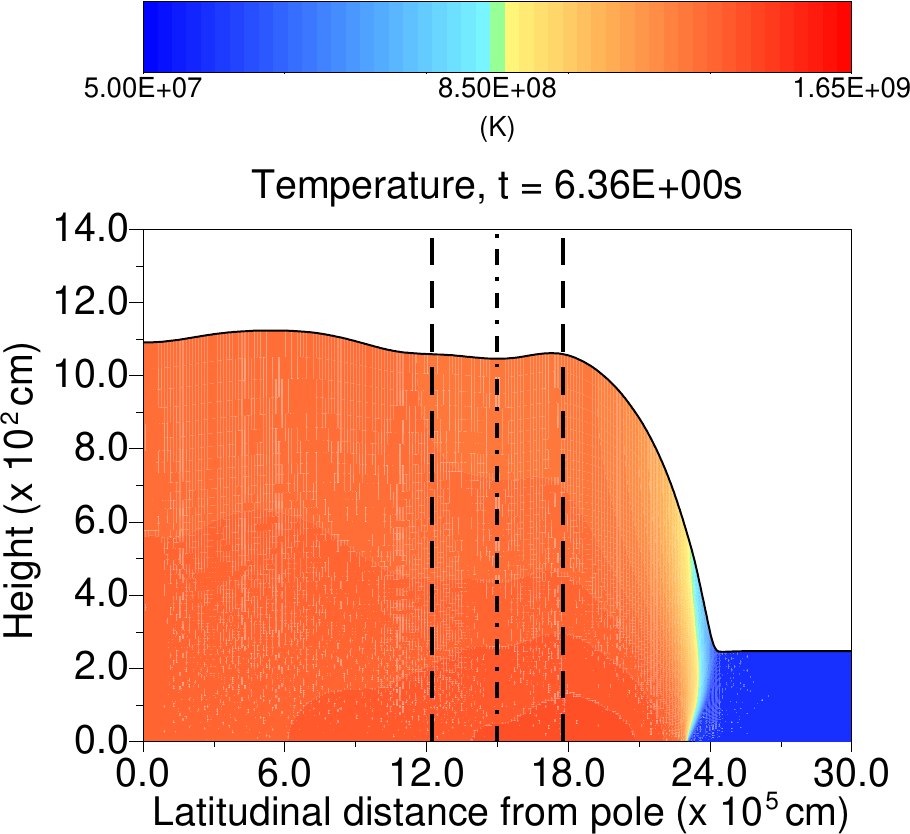}l)\\
  \caption[]{Continued. The crossing of the equator of P3: the flame
    burns with a horizontal front and eventually continues the
    propagation in the Southern hemisphere being confined again.}
  \label{fig:crossP3b}
\end{figure*}

\begin{figure*}
  \centering
  \includegraphics[width=0.45\textwidth]{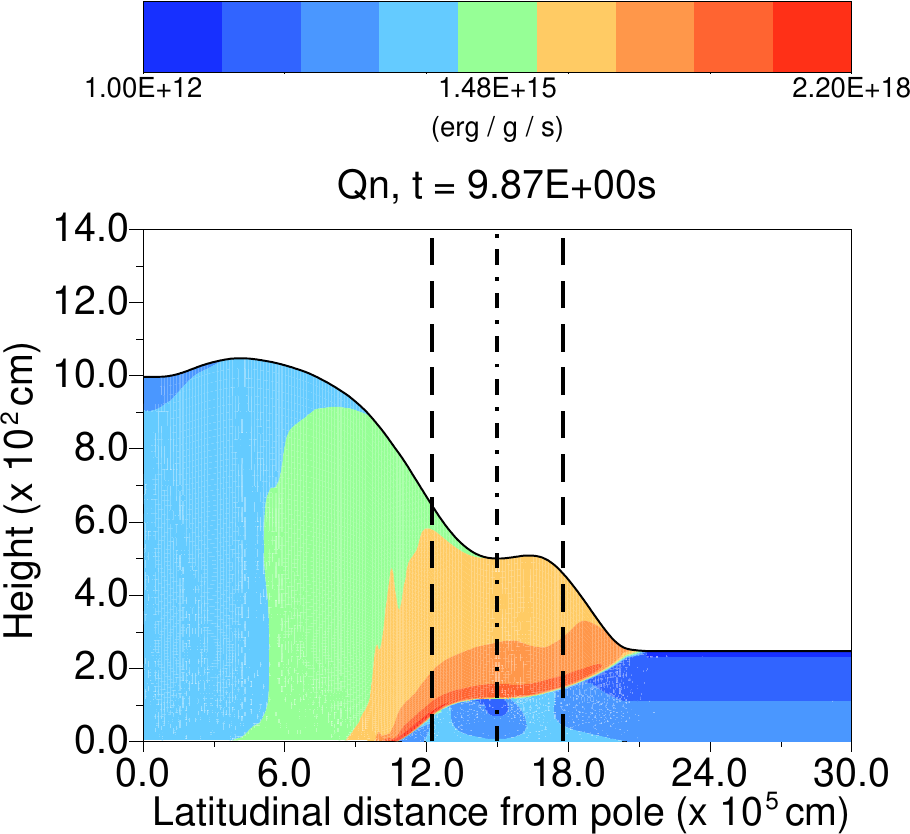}a)
  \hspace{\stretch{1}}
  \includegraphics[width=0.45\textwidth]{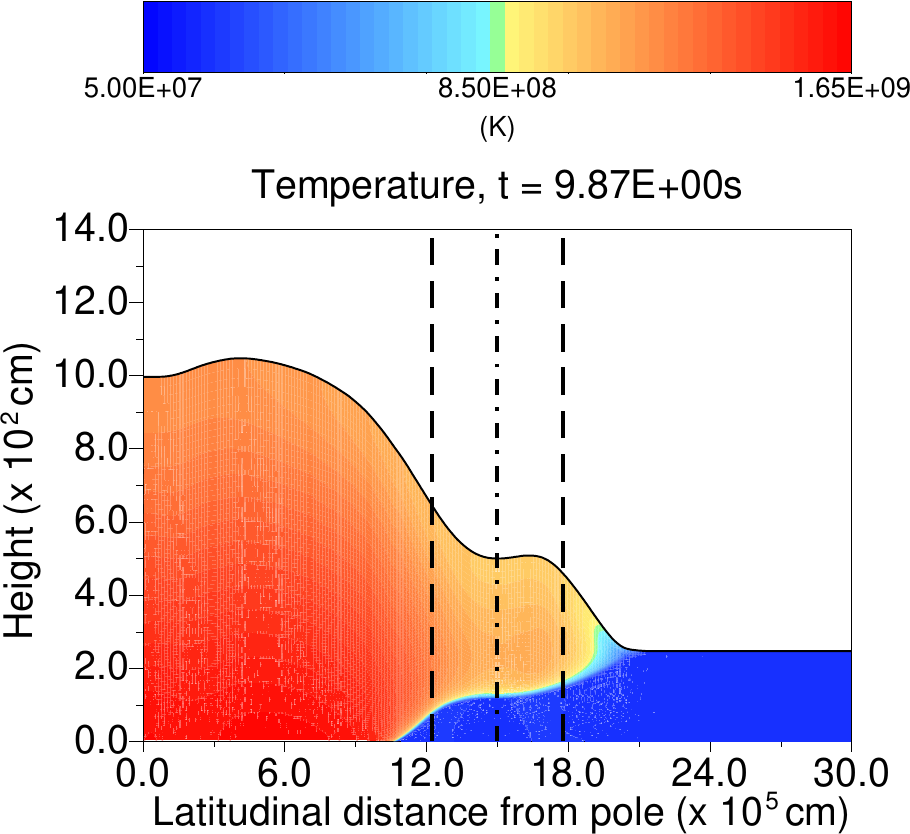}b)\\
  \includegraphics[width=0.45\textwidth]{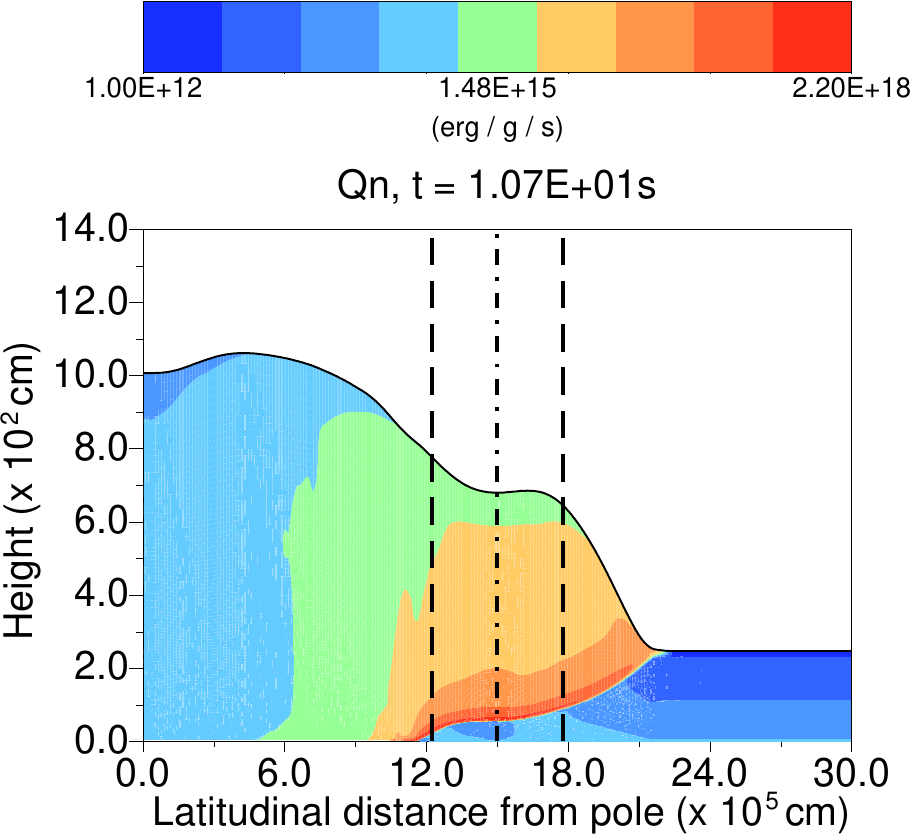}c)
  \hspace{\stretch{1}}
  \includegraphics[width=0.45\textwidth]{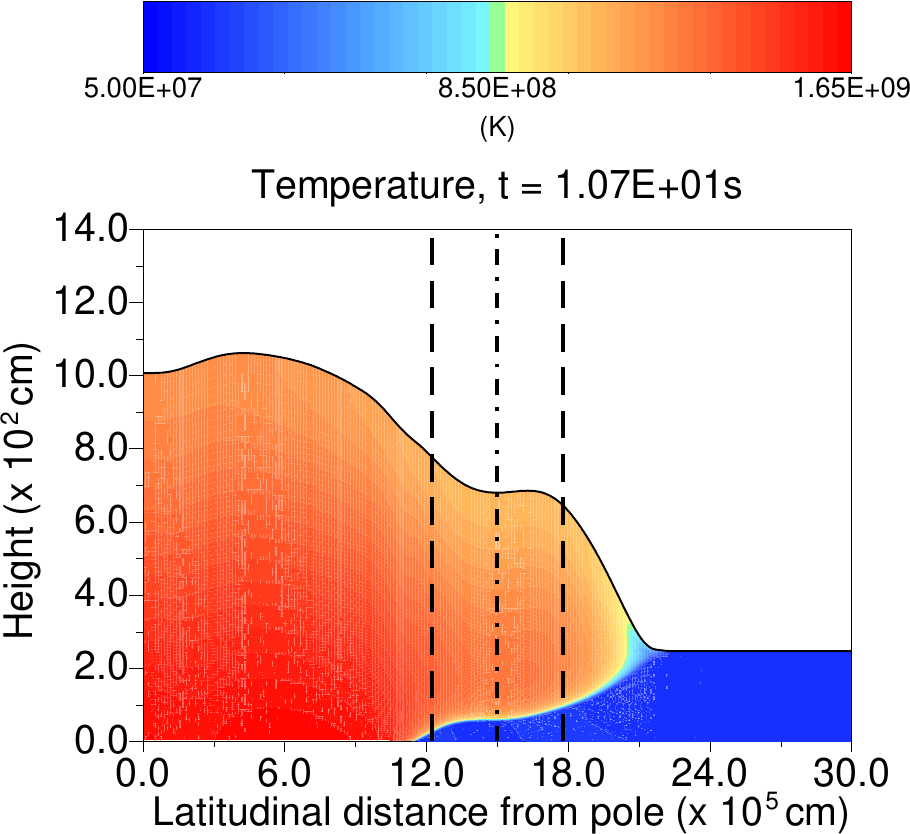}d)\\
  \includegraphics[width=0.45\textwidth]{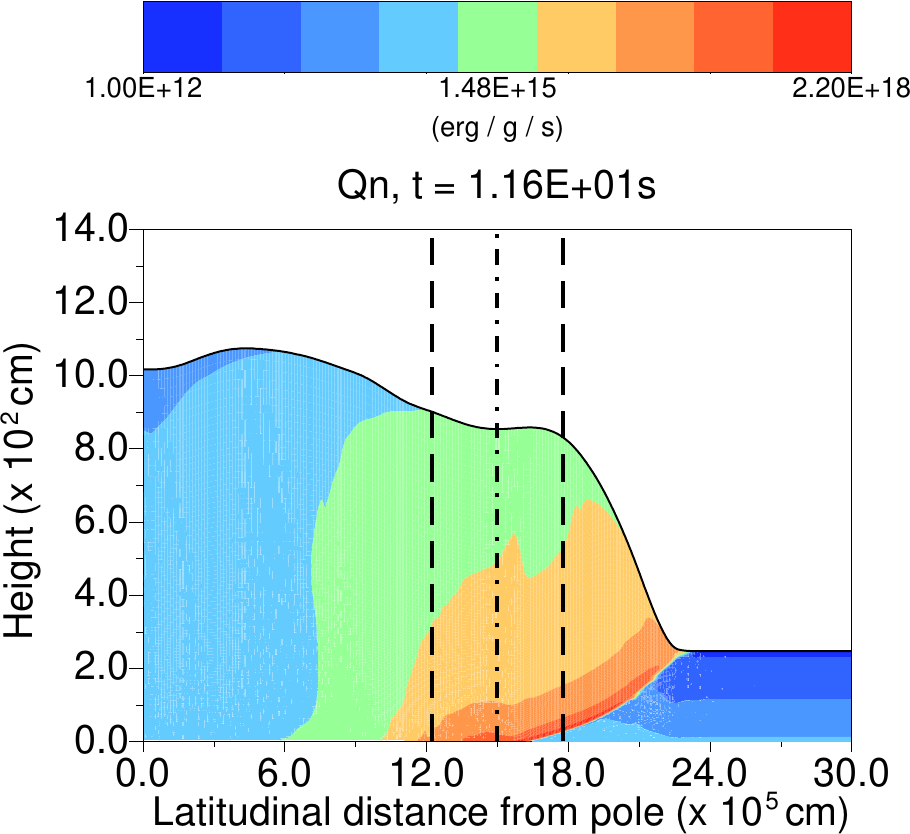}e)
  \hspace{\stretch{1}}
  \includegraphics[width=0.45\textwidth]{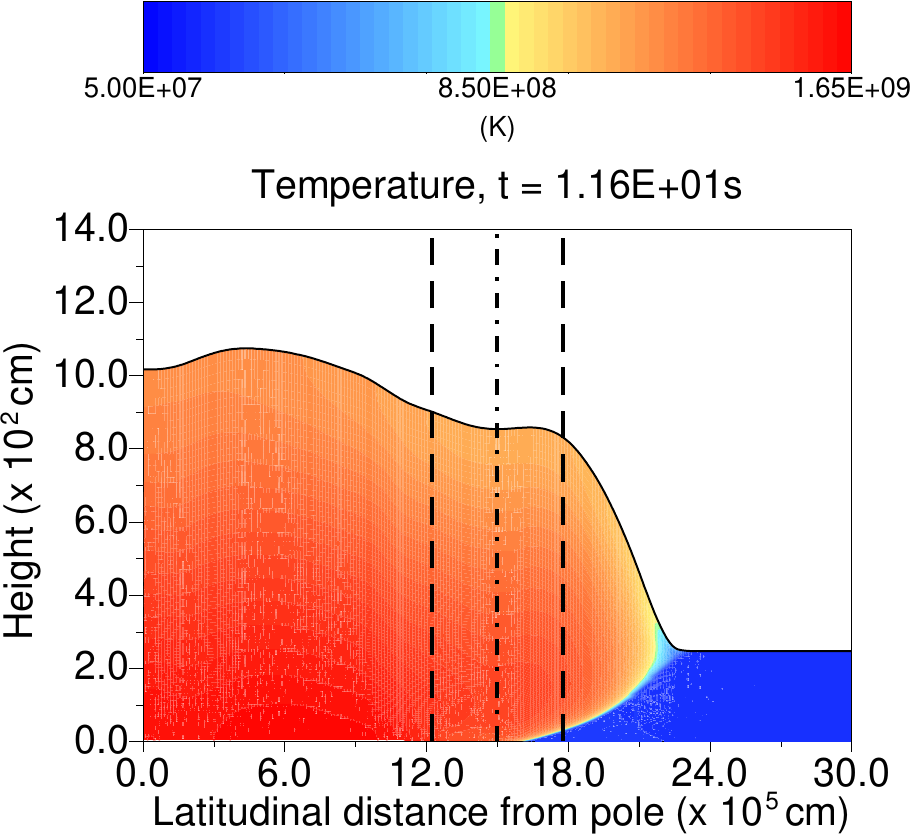}f)\\
  \caption{The crossing of equator of P5 ($\nu=450$ Hz, $\kaco=7$
    cm$^{2}$ g$^{-1}$). Same as \figref{fig:crossP3a}. The flame
    begins propagation in the Southern hemisphere before having
    reached the bottom of the simulation because of the higher
    opacity.}
  \label{fig:crossP5}
\end{figure*}

\begin{figure*}
  \centering
  \includegraphics[width=0.45\textwidth]{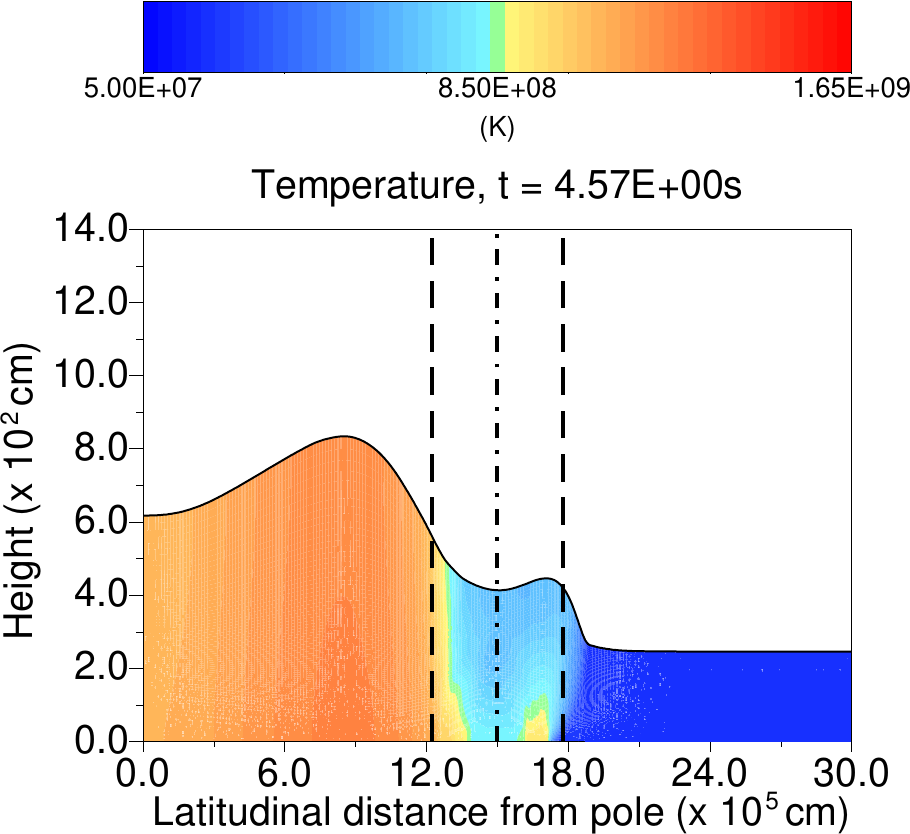}
  \hspace{\stretch{1}}
  \includegraphics[width=0.45\textwidth]{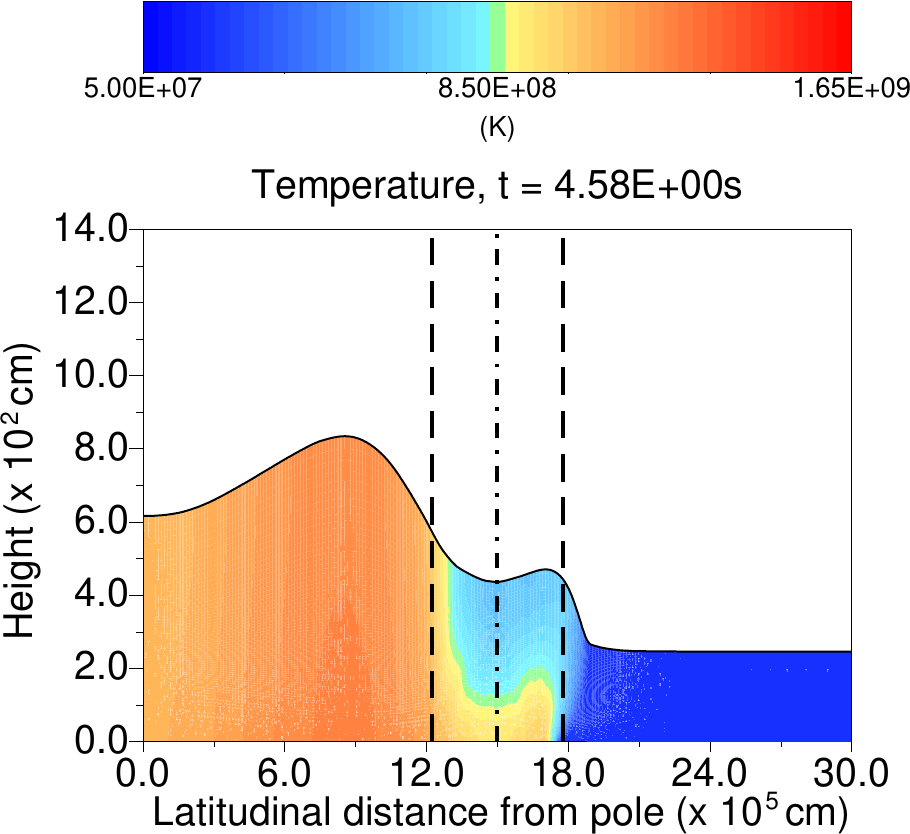}\\
  \caption{The crossing of equator of P7 ($\nu=450$ Hz,
    $\kaco=7\tent{-3}$ cm$^{2}$ g$^{-1}$). Only the temperature profile is
    plotted. Vertical lines indicate the equator (dash dotted) and the
    \belt{} (dashed, see \equlab{} \ref{eq:belt}).}
  \label{fig:crossP7}
\end{figure*}

\begin{figure*}
  \centering
  \includegraphics[width=0.45\textwidth]{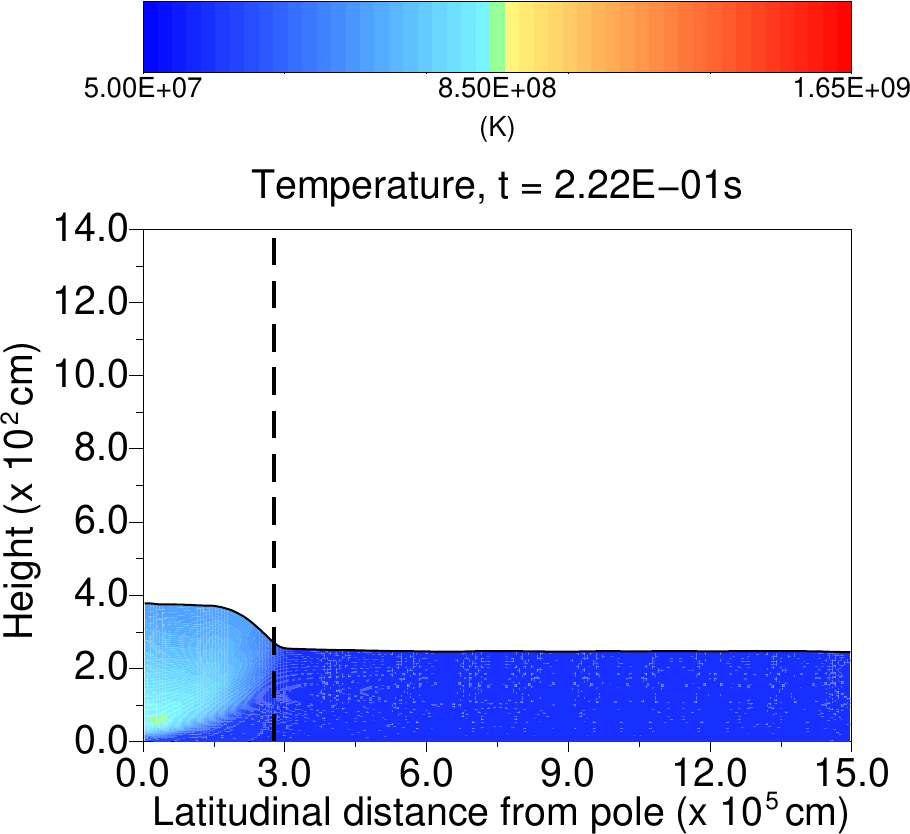}
  \hspace{\stretch{1}}
  \includegraphics[width=0.45\textwidth]{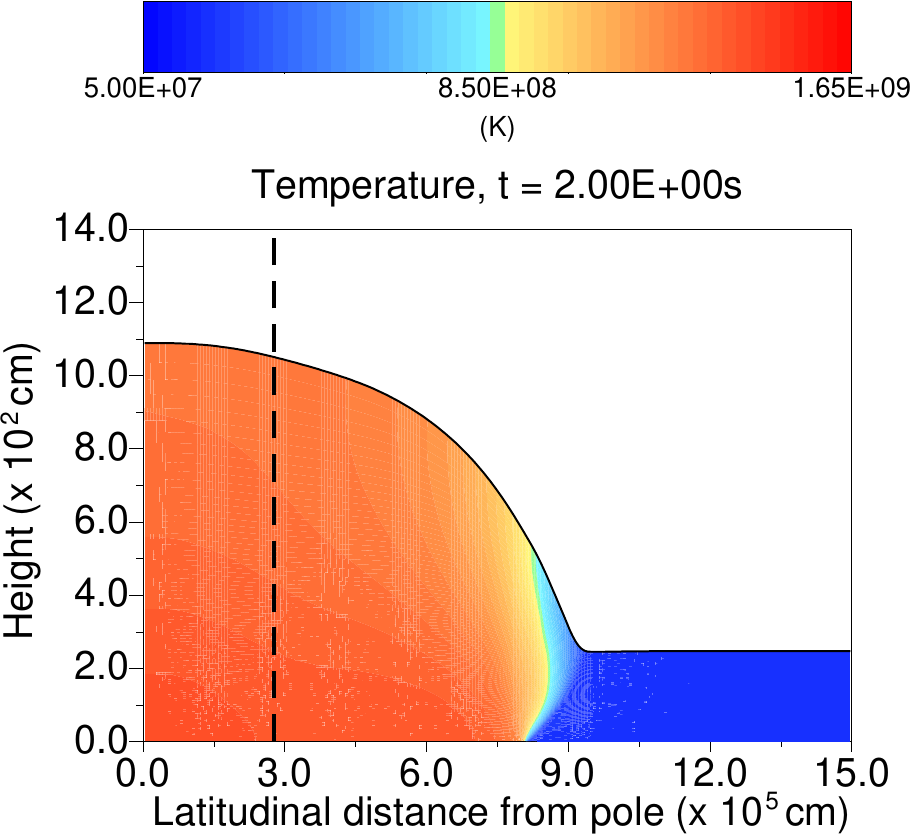}\\
  \caption[]{The ignition and propagation of the flame of E3. Same as
    \figref{fig:crossP7}. Apart from an initial transitional stage,
    the propagation is identical to the southern propagation of
    simulation P3.}
  \label{fig:crossE3}
\end{figure*}

The behaviour of simulations P2, $\nu=10^2$ Hz, and P4, $\nu=10^3$ Hz,
are qualitatively similar, the only difference being the variation in
the Coriolis force confinement with $1/\nu$ dependence on spin
frequency. Changing the thermal conductivity makes a greater
qualitative impact on the flame's dynamics, In \figref{fig:crossP5} we
show for comparison the crossing of the equator in run P5, where the
conductivity is much lower. As before, around the equator, the flame
front is almost horizontal and ignition propagates vertically (first
two panels of \figref{fig:crossP5}). However, in this case, the
vertical propagation is slow enough that the flame passes the
equatorial belt before it has reached the bottom of the
simulation. Note that temperature contours are more horizontal than in
the case of P3. On the other hand, \figref{fig:crossP7} shows the
behaviour of the flame when conductivity is higher, as in the case of
P7. Here also the flame stalls at the Northern boundary of the \belt{}
and a second flame ignites on the Southern boundary, but the second
flame first reaches the bottom of the simulation and only then merges
with the first one at the equator, before resuming the propagation
towards the south pole. Another important difference is that, in this
latter case, the north hemisphere has already cooled significantly
when the flame is passing the equator. When the flame has reached the
south pole, the temperature at the north pole is $\sim 7.7\tent{8}$ K
as opposed to $1.2\tent{9}$ K at the south pole. All these aspects can
be explained in terms of conduction timescales in the different
regimes. Finally, in \figref{fig:crossE3} we show for comparison what
happens when the ignition is at the equator, under the same conditions
of spin and conductivity, for the case of $\nu = 450$ Hz. Note how at
early stages the burning develops mostly within the \belt{} and, after
the initial transitional stage, the propagation is almost identical to
the second part of the simulation for polar ignition.

\subsection{Directionality of flame propagation}
\label{sec:propfit}

Here we will describe in more detail the features of the track
followed by the propagating flame as a function of time and position
on the star. In order to follow the propagation, we define the
position of the flame as the horizontal position, at each time, where
the burning rate is maximum.

\begin{figure}
  \includegraphics[width=0.45\textwidth]{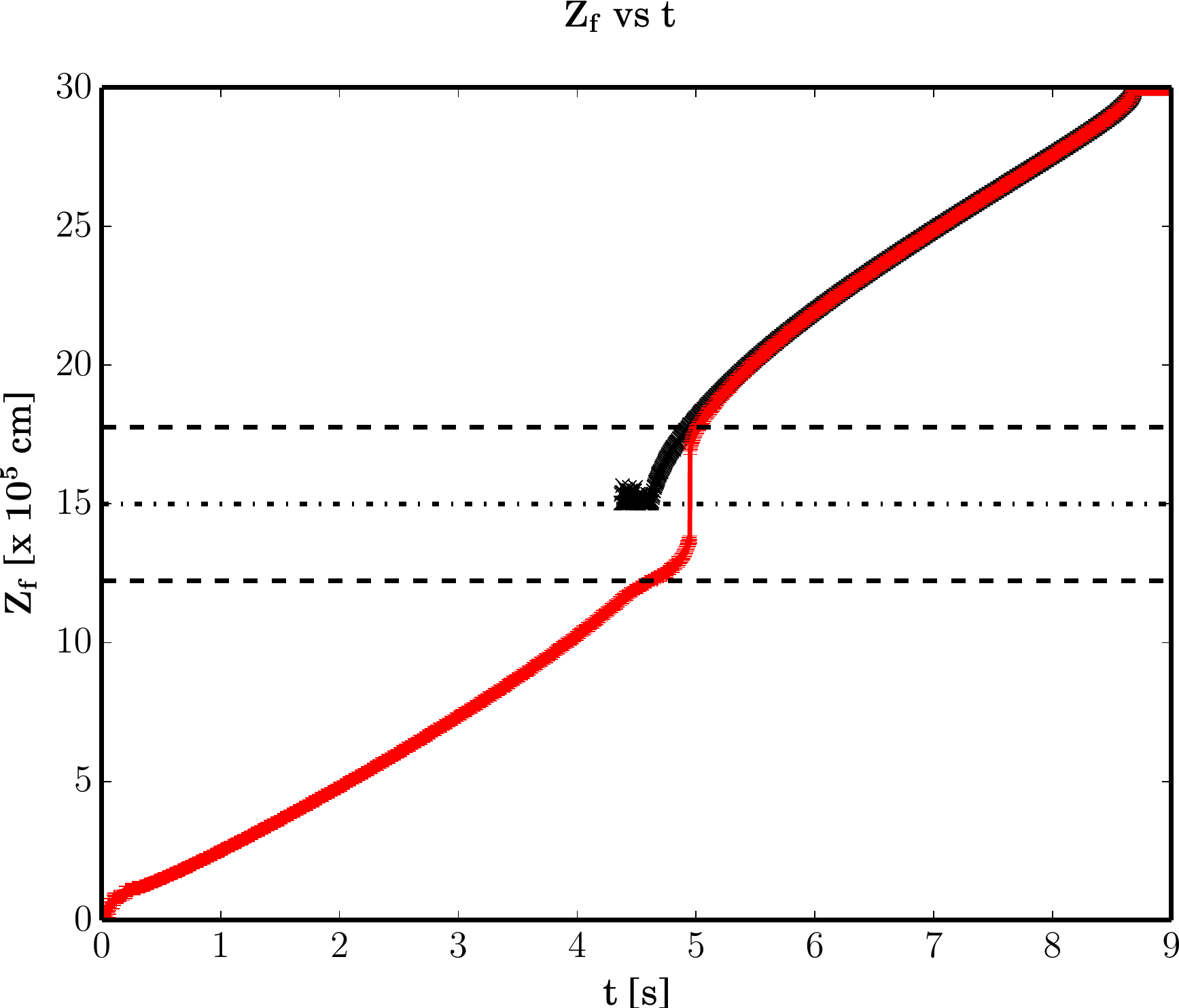}
  \caption{Horizontal position of the propagating front for simulation
    P3 (red) and E3 (black). Horizontal lines indicate the equator
    (dash dotted) and the belt as given by equation \eqr{eq:belt}. For
    run P3, the crossing of the equator is clearly visible as the
    almost vertical transition around $t = 5$ s. For run E3, the
    irregularities at the beginning correspond to the initial
    transient phase of flame ignition. The two simulations coincide to
    high degree, but the propagation from the equator to the south
    pole is not a mirror image of the propagation from the north pole
    to the equator.}
  \label{fig:propP3}
\end{figure}

\figref{fig:propP3} plots position versus time for the flame front for
our reference run P3 (in red) and for comparison shows the propagation
for the same parameters in the case of equatorial ignition (run E3, in
black). As for run P3, in \figref{fig:propP3}, near $t = 0$ s, there
is a transitional phase when the flame is starting, then the proper
propagation begins. At $t \sim 4.5$ s there is a noticeable decrease
in the speed. That feature is due to the effect of the \belt{} on the
flame structure. Indeed, the forward section of the front is inside
the \belt{} and the hot fluid is already slipping through it. The
temperature contours in panel (b) of \figref{fig:crossP3a} clearly show
the passage of the fluid. The missing heating contribution of the
slipping fluid is noticeable in the decrease of speed of the front.

When the hot fluid is within the \belt, the most vigorously burning
side is still the northern one, as can be seen in panel (e) of
\figref{fig:crossP3b} and in \figref{fig:propP3} until times $t
\lesssim 5$ s. At $t \sim 5$ s, the flat flame reaches the bottom and
starts propagating into the Southern hemisphere.  This can be
recognized in the sudden jump past the equator in
\figref{fig:propP3}. From then on, the flame continues again under the
effect of an increasing Coriolis force. This second part of the
propagation overlaps almost perfectly with that of the flame igniting
at the equator as can be seen in the figure, where the black crosses
are almost invisible below the red ones, apart from the initial
transitional stages of the ignition at the equator. Now we want to
draw attention to an unexpected fact: the propagation in the Northern
hemisphere is \emph{not} a mirror image of the propagation in the
Southern hemisphere.

Having verified that this was not a numerical effect\footnote{We
  performed a number of tests to verify that numerical effects and
  operation ordering effects could be ruled out: we changed the sign
  of the spin frequency for both equator and polar ignition and we
  changed the ignition position. We ignited at the south pole with
  propagation in both full and half domain and ignited at the equator
  propagating northwards. The conclusion is robust: every simulation
  igniting from a pole will cross the equator and then propagate to
  the other pole; in this second half of the propagation the flame is
  \emph{faster}. The propagation of a flame started at the equator, no
  matter towards which pole, coincides with this latter regime.}, we
proceeded to explore the physical cause of the phenomenon. In previous
runs, where the terms with $1/\tan\theta$ were not implemented yet, we
saw the same effect. In the production simulations, we find that the
asymmetry decreases for increasing spin and for increasing
conduction. Therefore, we think that the asymmetry may originate in
the balance between heat gains and losses and the asymmetry of
propagation regimes. At a given colatitude, the fluid propagating
towards the equator is more confined behind the front than it is in
front of it, while, when the flame propagates towards the pole,
confinement is smaller behind and higher in front of the flame. Higher
confinement at the front probably reduces heat losses via surface
cooling, speeding up the flame, while higher confinement in the back
prevents the hot fluid from contributing to the heating of the cold
fluid, slowing down the flame. The absolute value of the rate of
change of confinement (i.e. of the Rossby Radius) depends only on the
colatitude, but its sign depends on the direction of the propagation,
hence the asymmetry. Since this effect scales inversely with spin
frequency, this would explain the decreasing effect with increasing
spin.

This dependence of propagation speed on the direction of propagation
is an important fact that should be taken into account when simulating
flame propagation. Prompted by these considerations, we tried to fit
the propagation of the flame in both the hemispheres (see
\tabref{tab:fits}). If one assumes for the speed of the flame front
the $1/\nu$ dependence described in \cite{\ca} and
\cite{art-2002-spit-levin-ush}, then $v_{\rm{flame}} = \dot z \propto
1 / \cos \theta$, with $\theta = \z / \rstar$, and
\begin{equation}
  \dot \theta = \frac{\dot\theta_0}{\cos\theta}
\end{equation}
That leads to
\begin{equation}
\label{eq:line}
  \sin\theta = \dot\theta_0 t + I
\end{equation}
where $I$ is a constant of integration that takes into account the
fact that the fit does not start at $t = 0$ in our simulations.  The
results of the fit of a line to $\sin\theta$ versus time are reported
in \tabref{tab:fits}, for both cases of propagation towards the
equator and towards the pole; they are valid between $t_1$ and $t_2$.

However, as it should be expected, those fits were not very good:
instead we found that a law of the kind
\begin{equation}
\label{eq:parabola}
  \sin\theta = A t^3 + B t^2 + C t + D
\end{equation}
gives much better fits, as evaluated by the averaged weighted sum of
the residuals:
\begin{equation}
\label{eq:chidue}
\chi = \sqrt{\frac{1}{N-\np}\sum_1^{N}
(\frac{\sin\theta_{\rm i, fit} - 
    \sin\theta_{\rm i}}{\delta \sin\theta})^2}
\end{equation}
where $\delta \sin\theta= \cos\theta\delta\z/\rstar$ is the error in
the position of the flame given by propagating the error on the
position on the grid and $\np$ is the number of parameters
fitted. \tabref{tab:fits} reports the values for the fits. The time
between $t_2$ for the case going from pole to equator and $t_1$ when
going from equator to pole is approximately the `stalling' time at
the equator. These empirical fits could be used for simulating flame
propagation using a prescription of the type \eqr{eq:parabola} when
dealing with meridional propagation or at least, since in
  general different conditions of the ocean may affect the propagation
  time, they should give a measure of the asymmetry of the
propagation towards or away from the equator.

Finally, a remark on the equatorial crossing.  Looking at
\tabref{tab:fits} and comparing the values of $t_2$ for the P-E
section to those of $t_1$ for the E-P one, we have an idea of the
equatorial crossing time.  This ranges from $0.32$ to $2.64$ s,
depending on the effective opacity and, for our fiducial opacity
$\kaco=0.07$ cm$^{2}$ g$^{-1}$, it is on average $\sim 0.55$ s.

\begin{table*}
  \begin{center}
    \begin{tabular}{|c|c|r|r|r|r|r|r|r|r|r|r}
      \hline
      Type & Run & $t_1$ [s] & $t_2$ [s] & $A$ [$10^{-2}$ s$^{-3}$] 
        & $B$ [$10^{-2}$ s$^{-2}$] & $C$ [$10^{-1}$ s$^{-1}$] 
        & $D$ [] & $\chi_3$ & $\dot\theta_0$ [$10^{-1}$ s$^{-1}$] 
        & $I$ [] & $\chi_1$ \\
      \hline
      P--E & 2 & $0.50$ & $2.04$ & $-19.56$ & $94.31$ &  $-9.05$ & $0.32$ & $ 0.36$ &  $4.73$ & $-0.27$ & $ 3.21$ \\
      P--E & 3 & $0.26$ & $4.39$ & $ -0.42$ & $ 2.26$ &  $ 1.82$ & $0.06$ & $ 0.51$ &  $2.03$ & $ 0.07$ & $ 1.02$ \\
      P--E & 4 & $0.22$ & $6.09$ & $ -0.12$ & $ 0.34$ &  $ 1.68$ & $0.08$ & $ 0.18$ &  $1.43$ & $ 0.14$ & $ 1.38$ \\
      P--E & 5 & $0.55$ & $9.26$ & $ -0.09$ & $ 1.21$ &  $ 0.42$ & $0.11$ & $ 0.29$ &  $0.89$ & $ 0.07$ & $ 1.51$ \\
      P--E & 6 & $0.37$ & $6.30$ & $ -0.26$ & $ 2.60$ &  $ 0.72$ & $0.08$ & $ 0.17$ &  $1.42$ & $ 0.04$ & $ 2.51$ \\
      P--E & 7 & $0.14$ & $4.25$ & $ -0.03$ & $-0.65$ &  $ 2.43$ & $0.07$ & $ 0.20$ &  $2.07$ & $ 0.10$ & $ 2.81$ \\\hline
      E--P & 2 & $2.59$ & $3.51$ & $-56.51$ &$506.48$ &$-158.41$ &$17.71$ & $ 0.67$ & $-7.85$ & $ 2.87$ & $ 1.32$ \\
      E--P & 3 & $4.96$ & $8.70$ & $ -0.15$ & $ 1.63$ &  $-2.59$ & $2.04$ & $ 0.77$ & $-2.37$ & $ 2.16$ & $ 1.97$ \\
      E--P & 4 & $6.63$ &$12.44$ & $  0.04$ & $-2.09$ &  $ 1.12$ & $1.04$ & $ 0.32$ & $-1.56$ & $ 2.05$ & $ 1.12$ \\
      E--P & 5 &$11.90$ &$18.80$ & $ -0.06$ & $ 2.24$ &  $-3.92$ & $3.46$ & $ 0.54$ & $-1.24$ & $ 2.45$ & $ 2.14$ \\
      E--P & 6 & $7.44$ &$12.50$ & $ -0.12$ & $ 2.51$ &  $-3.24$ & $2.48$ & $ 0.25$ & $-1.71$ & $ 2.27$ & $ 1.22$ \\
      E--P & 7 & $4.57$ & $8.43$ & $ -0.33$ & $ 5.72$ &  $-5.57$ & $2.64$ & $ 0.34$ & $-2.37$ & $ 2.06$ & $ 0.85$ \\
      \hline
    \end{tabular}

    \caption{Numerical parameters for the fits to the flame position
      during propagation. $A$, $B$ $C$ and $D$ should be used for
      \eqr{eq:parabola}, while $\dot\theta_0$ and $I$ apply to
      \eqr{eq:line}. The first values are for flames going from pole
      to equator (P-E), the second ones from equator to pole
      (E-P). The parameters of the simulations can be read in
      \tabref{tab:poleq} with the corresponding number to those
      reported in the second column.  Also reported are the $\chi$
      values as from \eqr{eq:chidue} for the cases of linear $\chi_1$
      and cubic $\chi_3$ interpolation.}

    \label{tab:fits}
  \end{center}
\end{table*}

\subsection{Flame on slowly rotating NSs}
\label{sec:norot}

From previous studies, it was clear that the Coriolis force is
important for flame propagation, but there exist cases, like \tfive{}
spinning at $\nu=11$ Hz \citep[see][]{atel-2010-alta-etal,
  art-2011-cavecchi-etal} where the rotation cannot provide
confinement; nonetheless, they show pulsations during type I
bursts. We therefore studied cases of low rotation.

We simulated a non rotating star, even though our initial conditions
are not strictly speaking appropriate for this case, since there is no
Coriolis force to confine the initial hot fluid. In this simulation
the fluid spreads over the entire surface almost instantaneously and
eventually burns, after $\sim 30$ s, in what is practically a 1D
configuration. However, since most if not all NSs rotate we
do not discuss this simulation any further.

\begin{figure*}
  \centering
  \includegraphics[width=0.45\textwidth]{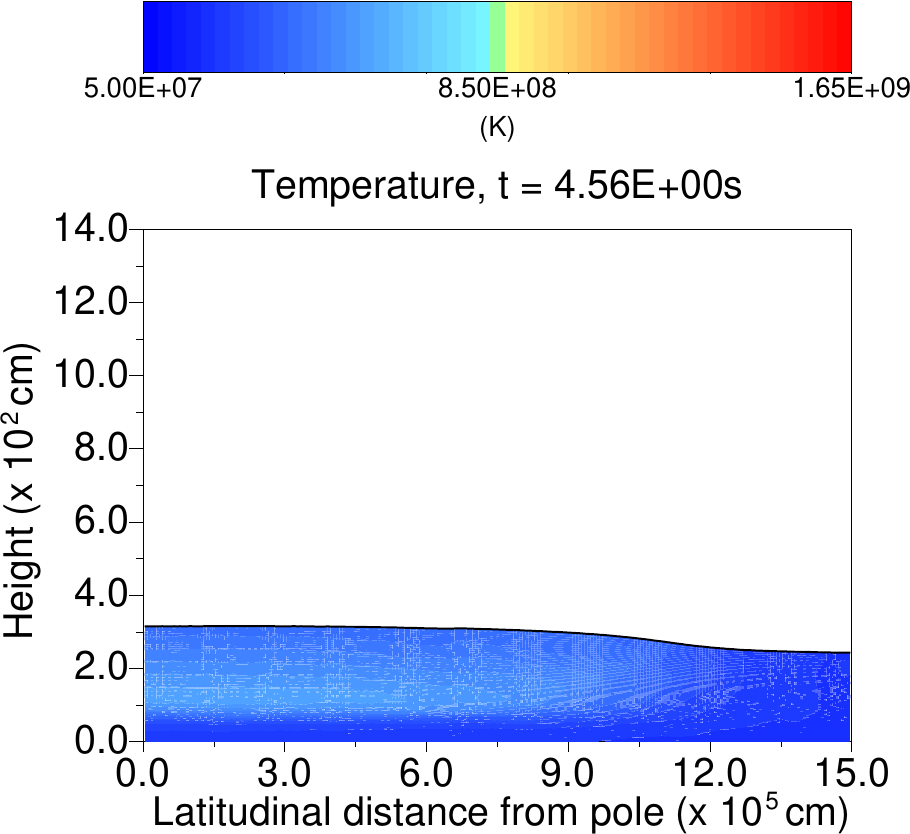}
  \hspace{\stretch{1}}
  \includegraphics[width=0.45\textwidth]{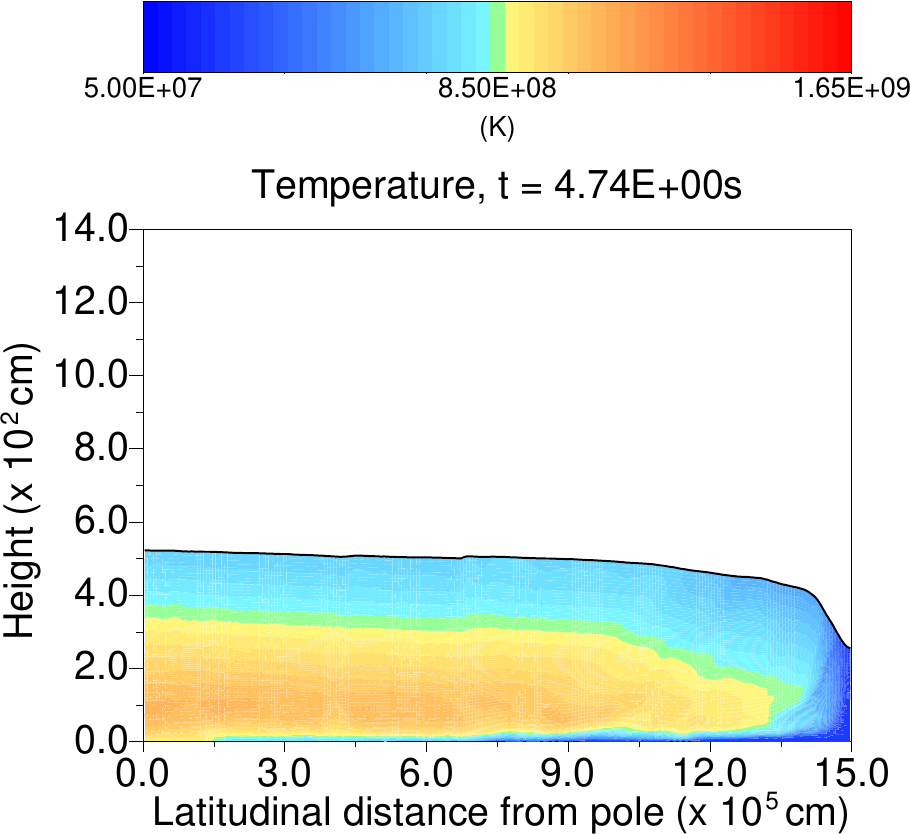}\\
  \caption{The late ignition of simulation E1 ($\nu = 10$ Hz).}
  \label{fig:E1nohotspot}
\end{figure*}

We then considered a case with $\nu = 10$ Hz, comparable to the
frequency of \tfive{}, for both polar and equatorial ignition. We
found that, on one hand, after the fluid has oscillated a few times,
simulation E1 ignites at $t \sim 4.5$ s. The temperature quickly
exceeds $10^{9}$ K, starting the runaway, and the flame is visibly
burning almost the whole domain, not being substantially confined (see
\figref{fig:E1nohotspot}) similarly to the regimes of
self-ignition. On the other hand, simulation P1, after a similar
sloshing, has not yet ignited significantly after $t \sim 28$ s, but
the temperature has been increasing at the pole, where $T \sim 10^{9}$
K, while most of the fluid is still cold. Since the temperature is
increasing and fluid is burning at the pole, it is possible that at
later times the burning could become significant, but we do not count
this as a flame ignited at the initial `hot-spot' and then
propagated. Indeed, in the case of polar ignition the fluid that is
hot at the pole is slowly heating the rest, but spreading over most of
the surface. What we see confirms the fact that a sufficient amount of
matter needs to be confined for ignition to happen. The difference in
behaviour between simulation E1 and P1 is probably due to the
difference in the extent of the two simulations.

\section{Discussion and conclusions}
\label{sec:conc}

In \citet{\ca}, we showed the importance of the Coriolis force
confinement for flame propagation considering only cases of spreading
under constant Coriolis parameter. However, we did not consider
meridional propagation, where Coriolis effect diminishes from pole to
equator, and a question arose about whether the flame could cross the
equator, where the confinement is absent. Moreover, we did not
consider cases where the spin frequency was incapable of providing
confinement at any latitude on the star. We analysed these problems in
this paper.

First of all, we reported on our simulations that show how a flame
ignited at one pole can successfully reach the equator and cross it to
proceed to the other hemisphere, at least for $\nu \ge 10^2$ Hz. We
also showed how thermal conductivity can affect the propagation and
the general temperature profile over the surface. Finally, we showed
that in our simulations there is a difference in the propagation of
the flame from pole to equator and from equator to pole.

Equatorial crossing implies that the full star is probably burning
during type I bursts, as is usually assumed. However, depending
  on the conductivity and the details of heat transport, one
  hemisphere may be significantly cooler than the other. For example,
  in the case of simulation P7 (\secref{sec:poligni} and
  \figref{fig:crossP7}), where the conductivity is high enough, the
  highest temperature at the south pole is $\sim 1.6$ times the
  coolest one at the north pole, when propagation has finished. The
  thermal time scale in the vertical direction above the burning layer
  is, conservatively, $\lesssim 1$ s
  \citep{art-2000-cum-bild,art-2006-wein-bild-scha}, while in the
  horizontal direction it is longer by a factor approximately given by
  the square of the ratio of the length scales $(\rstar/H)^2 \sim 10^6
  - 10^4$, where $H$ is the thickness of the fluid above the burning
  layer and $\rstar$ is the star radius. Since the propagation takes
  up to few seconds, the difference in temperature at the bottom
  should be reflected in the emitting layers of the photosphere. This
should be taken into account when analysing light curves which fit
only one temperature. The temperature derived would be an average of
the surface distribution and, for example, could affect conclusions
about NS radius.

We found that the flame takes up to a few seconds to cross the
equator, and, for realistic values of the opacity ($\kaco
  \approx 0.07$ cm$^{2}$ g$^{-1}$, our fiducial value), the time
needed decreases below 1s; of the order the time it takes
  to the flame to propagate downwards. This result has implications
for all models and interpretations that have invoked any form of
``stalling'' of the flame at the equator. For example, the values we
measure are too short compared to the times that
\cite{art-2006-bhatta-stro} needed to explain double peak bursts,
which are of the order of a few seconds. Those authors required the flame
propagation to stop at the equator in order to explain bursts with
double peaks: our simulations show that hydrodynamics alone does not
allow for sufficient stalling. Of course, other mechanisms to stall
the flame are still possible: in particular the role of magnetic field
has to be considered carefully; as must the important case when the
flame ignites at mid latitudes, so that there is not a single ring of
fire propagating from pole to pole: this could imply that less burning
fluid reaches the equatorial band, leading to a possible flume out. We
plan to address this problem in a subsequent paper.

The asymmetry we found in the propagation from pole to equator as
compared to from equator to pole, led us to provide a very basic
fitting formula that could be used in order to simulate the
propagation of a flame in a parametrized way, or at least provide a
measure of the asymmetry between the two regimes. In particular, one
can derive how much faster propagation from equator to pole is with
respect to propagation from pole to equator. Note for example that the
papers of \cite{art-2006-bhatta-stro} and, more recently,
\cite{art-2014-chako-bhatta} assumed that the velocity of the flame
depends only on the latitude and not also on the direction.

Finally, it is interesting to consider the simulations at frequency
$\nu = 10$ Hz. In such simulations the Coriolis force was not strong
enough to confine the hot fluid. However, the fluid did eventually
ignite, albeit on a much longer timescale and the flame and front were
significantly different in nature with respect to the other, confined,
cases: a great fraction of the fluid ignited almost simultaneously and
only propagated through a small distance, similar to the regime of
self ignition. The time needed for local ignition to finally
happen probably depends on the interplay between the small confinement
provided by the weak Coriolis force, the extent of the domain, as
evidenced by the difference between the equatorial ignition and the
polar ignition, the cooling prescription and the burning rate and
energy release. However, one more conclusion can be drawn: our
simulations support the arguments used by
\cite{art-2011-cavecchi-etal}, who suggested that in the pulsar
\tfive{} spinning at $11$ Hz, the presence of a hot-spot could not be
achieved by the Coriolis force effects\footnote{The fact that in our
  simulations the pole is hotter could not be considered a reason for
  any pulsation, since, being exactly at the pole, it is rotationally
  symmetric.} and therefore proposed that the surface asymmetries
responsible for the strong measured pulsations might be caused by
magnetic confinement.

{\it Acknowledgements.} We thank Frank Timmes for making his
  astrophysical routines publicly available. We also thank an
  anonymous referee for useful comments that improved this
  manuscript. YC and AW acknowledge support from an NWO Vrije
Competitie grant ref. no. 614.001.201 (P.I. Watts). This paper
benefitted from NASA's Astrophysics Data System.

\bibliographystyle{mn2e}
\bibliography{ms-bibfile}

\label{lastpage}
\end{document}